\documentclass[12pt, draftclsnofoot, onecolumn]{IEEEtran}

\usepackage[ruled,vlined]{algorithm2e}
\SetArgSty{textup}
\usepackage{amsmath,nccmath}
\usepackage{amssymb}
\usepackage{array}
\usepackage{bbm}
\usepackage{cite}
\usepackage{color}
\usepackage{float}
\usepackage{mathtools}
\usepackage{multirow}
\usepackage{setspace}
\usepackage[caption=false]{subfig}
\newcolumntype{M}[1]{>{\centering\arraybackslash}m{#1}}
\newcolumntype{P}[1]{>{\centering\arraybackslash}p{#1}}
\newenvironment{subroutine}[1][htb]{\begin{algorithm}[#1]}{\end{algorithm}}

\newtheorem{remark}{Remark}
\makeatletter
\let\sv@thm\@thm
\def\@thm{\let\indent\relax\sv@thm}
\makeatother
\restylefloat{table}

\begin{document}
\title{Over-the-Air Design of GAN Training for mmWave MIMO Channel Estimation}
\author{Akash Doshi, Manan Gupta and Jeffrey G. Andrews \thanks{The authors are with 6G@UT in the Wireless Networking and Communications Group and the Dept. of Electrical and Computer Engineering at the University of Texas at Austin, TX 78712 (e-mail: akashsdoshi, g.manan@utexas.edu, jandrews@ece.utexas.edu). Date of current version: May 22, 2022. This work was supported by NVIDIA, Qualcomm Innovation Fellowship (QIF) and the NSF under Grants CNS-2148141 and CCF-2008710.} }

\maketitle 
\normalsize
\begin{abstract}
Future wireless systems are trending towards higher carrier frequencies that offer larger communication bandwidth but necessitate the use of large antenna arrays. Existing signal processing techniques for channel estimation do not scale well to this ``high-dimensional" regime in terms of performance and pilot overhead. Meanwhile, training deep learning based approaches for channel estimation requires large labeled datasets mapping pilot measurements to clean channel realizations, which can only be generated offline using simulated channels. In this paper, we develop a novel unsupervised over-the-air (OTA) algorithm that utilizes noisy received pilot measurements to train a deep generative model to output beamspace MIMO channel realizations. Our approach leverages Generative Adversarial Networks (GAN), while using a conditional input to distinguish between Line-of-Sight (LOS) and Non-Line-of-Sight (NLOS) channel realizations. We also present a federated implementation of the OTA algorithm that distributes the GAN training over multiple users and greatly reduces the user side computation. We then formulate channel estimation from a limited number of pilot measurements as an inverse problem and reconstruct the channel by optimizing the input vector of the trained generative model. Our proposed approach significantly outperforms Orthogonal Matching Pursuit on both LOS and NLOS channel models, and EM-GM-AMP -- an Approximate Message Passing algorithm -- on LOS channel models, while achieving comparable performance on NLOS channel models in terms of the normalized channel reconstruction error. More importantly, our proposed framework has the potential to be trained online using real noisy pilot measurements, is not restricted to a specific channel model and can even be utilized for a federated OTA design of a dataset generator from noisy data.
\end{abstract}


\newpage 

\section{Introduction} \label{sec:intro}
\subsection{Motivation}
Channel estimation (CE) in 6G and beyond will increasingly be performed for larger antenna arrays at both the base station (BS) and user (UE), increasing the dimensionality and complexity of the problem, as future communication systems tend towards progressively higher carrier frequencies in the mmWave and sub-THz range \cite{rappaport2019wireless, elayan2018terahertz}. While channel estimation at sub-6 GHz in practical wireless deployments is usually performed using Least Squares (LS) or minimum mean squared error (MMSE) estimators \cite{bjornson2017massive}, these do not scale to higher carrier frequencies which require large antenna arrays.

At higher carrier frequencies, the beamspace sparsity of MIMO channels is exploited to perform beam alignment -- steering the antenna arrays at the BS and UE to maximize SNR. Due to latency constraints, the number of candidate beams used to perform beam alignment is relatively small, limiting the SNR achieved. Consequently, researchers have been actively investigating learning-based approaches with an eye to improving both CE and beam alignment in this ``high-dimensional" regime. A common drawback of such attempts has been the requirement of large datasets of clean channel realizations for training. Such training datasets are either generated using statistical channel models or through ray-tracing for a given physical environment. What is desirable is a learning-based CE or beam alignment algorithm that can be trained using received (noisy) pilot measurements alone. This will eliminate performance losses stemming from inaccurate channel modelling and allow for online updates to the learned models. 

Deep generative modelling \cite{radford2015unsupervised} is a powerful unsupervised learning tool that can be used to implicitly learn complex probability distributions. Recently, it has also been utilized for solving \textit{inverse problems} i.e. reconstructing an unknown signal or image from observations \cite{bora2017compressed}. The forward process mapping the signal to the observation is typically non-invertible, hence it is impossible to uniquely reconstruct the signal in the absence of some prior knowledge \cite{ongie2020deep}. A trained deep generative network can be used to encode such a prior. In this paper, we describe a novel over-the-air approach based on deep generative modelling to learn the high-dimensional channel distribution from noisy pilot measurements and subsequently use the trained generative networks to accurately reconstruct the beamspace representation of the channel from compressive pilot measurements.

\subsection{Related Work} \label{subsec:related}
Channel measurements indicate that mmWave MIMO channels are sparse in their beamspace representation \cite{eliasi2017low} due to clustering of the paths into small, relatively narrowbeam clusters. To exploit this sparsity (that LS and MMSE cannot), several papers have provided a sparse formulation of mmWave CE in both the angular \cite{bajwa2010compressed} and delay-Doppler domain \cite{bajwa2009sparse} and subsequently employed Orthogonal Matching Pursuit (OMP) \cite{alkhateeb2014channel} and Basis Pursuit Denoising \cite{mendez2016hybrid} for sparse channel reconstruction. Approximate Message Passing (AMP) \cite{rangan2011generalized} and its variants such as EM-GM-AMP \cite{vila2013expectation} and VAMP \cite{rangan2019vector} have also proved to be robust compressed sensing (CS)-based approaches to sparse channel estimation \cite{wen2014channel, sun2018joint}.

CS-based approaches suffer from the drawback of assuming exact sparsity in the DFT basis, and result in poor channel reconstruction if the cross-correlation in the columns of the sensing matrix is high \cite{duarte2011structured}. Meanwhile deep learning based approaches to channel estimation have been making rapid advancements. A plethora of papers recently \cite{yang2019deep,He18Li,ru2019model,Dong19Gaspar,chun2019deep} have designed supervised deep learning techniques that provide pilot measurements as input to a neural network (NN) that outputs the desired channel. These suffer from the drawback of requiring a labeled dataset to train, and hence have to be trained offline. To circumvent the requirement of a labeled dataset, we designed an unsupervised CE algorithm \cite{balevi2020high, doshi2020compressed} that trained a deep generative network to output channel realizations using Generative Adversarial Networks (GAN) \cite{goodfellow2014generative}. Our proposed algorithm used the trained generative network for high-dimensional CE from a small number of pilots and provided a compressed representation of the estimated channel which could be used for channel reconstruction at the BS. By employing posterior sampling via Langevin dynamics in combination with a generative prior \cite{jalal2021robust} for channel estimation, \cite{arvinte2021deep} has further improved upon the performance of \cite{balevi2020high}, albeit at the cost of a significantly increased execution time which could run contrary to the block fading assumption made in their exposition. The generative CE algorithm of \cite{balevi2020high} was also applied to wideband channel estimation in \cite{balevi2021wideband}, at the cost of designing a separate generative model for each transmit and receive antenna pair, thus not exploiting spatial correlation.

However, the aforementioned generative channel estimation techniques and other works performing generative channel modelling \cite{orekondy2022mimo} require a large dataset of clean channel realizations to train a GAN. In addition, our prior work \cite{balevi2020high} has the following two limitations: i) Antenna spacing of $\lambda_c/10$ (where $\lambda_c = c/f_c$ and $f_c$ is the carrier frequency) in place of the conventional $\lambda_c/2$ was assumed to generate high spatial correlation in channel realizations and ii) the GAN was trained and its performance evaluated only on channels with a strong LOS component.

\subsection{Contributions}
We propose a design for an over-the-air (OTA) implementation of GAN training and generative channel estimation. We utilize received pilot measurements, instead of clean channel realizations, to train a Conditional Wasserstein GAN with no assumptions on the channel model or line-of-sight conditions. We propose several new ideas that while collectively overcoming the key real-world limitations of the initial work done in \cite{balevi2020high}, also set forth a new distributed design strategy -- FedPilotGAN -- as elaborated in our contributions below.

\textbf{Beamspace Generative Channel Estimation:} In order to relax the antenna spacing constraint from $\lambda_c/10$ to the conventional $\lambda_c/2$, we propose to solve generative channel estimation in the beamspace domain. This helps to exploit the angular clustering in mmWave MIMO channels, while simultaneously not imposing any constraints on the sparsity of the beamspace representation as is the case in traditional CS algorithms. Moreover, we demonstrate the advantage of using deep generative models over CS based approaches for solving inverse problems by showing that a trained generative prior eliminates the need for careful tuning of the sensing matrices.

\textbf{Training Wasserstein GANs for LOS and NLOS channels:} In \cite{balevi2020high}, we trained a Wasserstein GAN (WGAN) on channels with a dominating Line-of-Sight (LOS) component. By training the generator to output beamspace channel representations and improving the performance of WGAN using Gradient Penalty (GP) \cite{gulrajani2017improved}, we train generative models for a range of LOS and Non-Line-of-Sight (NLOS) MIMO channels with varying degrees of approximate sparsity in their beamspace representation. Subsequently, we train a single Conditional Wasserstein GAN (CWGAN) that takes as conditional input the channel's LOS/NLOS state to output samples from the entire aforementioned range of channels.

\textbf{Training a GAN from noisy pilot measurements:} In order to overcome the requirement of a dataset of clean channel realizations, we present a novel GAN architecture -- \textit{Pilot GAN} -- that combines LS channel estimation with Ambient GAN \cite{bora2018ambientgan} to train a generative model to output MIMO channel realizations given a dataset of full-rank noisy pilot measurements in a hybrid mmWave architecture. We also provide insights into the range of measurement models under which the Pilot GAN framework is applicable.

\textbf{Federated Pilot GAN:} We describe a federated implementation of Pilot GAN -- \textit{FedPilotGAN} -- to facilitate distributed GAN training from noisy pilot measurements over multiple users. This helps to reduce the UE side computation and move the computationally intensive generator training to the BS without impacting the performance of generative channel estimation, as will be demonstrated in Section \ref{subsec:results_pilot_gan}. We also highlight the ability of FedPilotGAN to design a dataset generator at a server (BS) using only noisy realizations collected at multiple users -- without transmitting data realizations between the server and the users.

\textbf{Pilot Conditional GAN:} We combine the aforementioned Pilot GAN with the CWGAN architecture to train a conditional generative model from noisy pilot measurements. Furthermore, we train a neural network (NN) to determine the LOS/NLOS state of the channel based on pilot measurements, and use its output as input to the conditional generator, thus designing an OTA GAN training algorithm based on pilot measurements. The resulting trained conditional generator is then used to perform CE from compressive pilot measurements across a variety of high-dimensional MIMO channels.

\subsection{Notation \& Organization}
We use bold uppercase $\mathbf{A}$ to denote a matrix and bold lowercase $\mathbf{a}$ to denote a vector. $||\mathbf{A}||_2$ is the Frobenius norm of $\mathbf{A}$ and $||\mathbf{a}||_0$ denotes the number of non-zero entries in $\mathbf{a}$. $\mathrm{diag}(\mathbf{A}_1, \ldots, \mathbf{A}_N)$ represents a block diagonal matrix whose diagonal entries are given by $\{\mathbf{A}_1, \ldots, \mathbf{A}_N\}$.  $\underline{\mathbf{A}}$ is a vector obtained by stacking the columns of $\mathbf{A}$. For matrix $\mathbf{A}$ of size $M \times N$ and matrix $\mathbf{B}$ of size $P \times Q$, $\mathbf{A} \otimes \mathbf{B}$ denotes the $MP \times NQ$ matrix of their Kronecker product. $\mathbb{E}[.]$ is the expectation operator, $\mathrm{sgn}(x) = 1$ if $x > 0$, $-1$ if $x < 0$ and $0$ otherwise and $\mathbb{P}(B)$ denotes the probability that an event $B$ occurs. $[a]$ denotes the set $\{1, \ldots, a\}$ for any integer $a$, while $[a] \times [b]$ denotes the Cartesian product of the two sets.

The paper is organized as follows. The system model is outlined in Section~\ref{sec:sys_model}, followed by a detailed description of the beamspace generative channel estimator in Section~\ref{sec:gce}. Each of the GAN architectures mentioned above is then expounded upon in Section~\ref{sec:gan_arch}, followed by the architectural and simulation details of these GANs in Section~\ref{sec:simulation}. The results are presented in Section~\ref{sec:results} and the paper concludes with final thoughts and future directions in Section~\ref{sec:conclusion}.

\section{System Model} \label{sec:sys_model}
Consider the problem of single user downlink (DL) narrowband mmWave MIMO channel estimation with a transmitter having $N_t$ antennas and the receiver having $N_r$ antennas. Denote the hybrid precoder at the transmitter by $\mathbf{F} \in \mathbb{C}^{N_t \times N_s}$, and the hybrid combiner at the receiver by $\mathbf{W} \in \mathbb{C}^{N_r \times N_s}$, with $N_s$ being the number of data streams that can be transmitted. 
With the MIMO channel denoted by $\mathbf{H} \in \mathbb{C}^{N_r \times N_t}$, the transmitted pilot symbols $\mathbf{S} \in \mathbb{C}^{N_s \times N_p}$ are received as
\begin{equation} \label{eq:sys_model}
    \mathbf{Y} = \mathbf{W}^H \mathbf{HFS} + \mathbf{W}^H\mathbf{N}, 
\end{equation}
where each element of $\mathbf{N} \in \mathbf{C}^{N_r \times N_p}$ is an independent and identically distributed (i.i.d.) complex Gaussian random variable with mean 0 and some variance $\sigma^2$. Implicitly assumed in \eqref{eq:sys_model} is a block fading model over $N_p$ pilot symbols i.e. a new i.i.d. channel realization $\mathbf{H}$ is chosen every $N_p$ time slots. In this paper, we assume a fully connected phase shifting network \cite{mendez2016hybrid}. We also constrain the angles realized by the phase shifters to quantized sets \cite{venugopal2017channel} given by 
\begin{equation}
    \mathcal{A} = \bigg\{0, \frac{2\pi}{2^{N_Q}}, \ldots, \frac{(2^{N_Q}-1)2\pi}{2^{N_Q}} \bigg\},
\end{equation}
where $N_Q$ is the number of quantization bits. We assume $N_{bit,t}$ and $N_{bit,r}$ phase shift quantization bits at the transmitter and receiver respectively, with the quantization sets denoted by $\mathcal{A}_t$ and $\mathcal{A}_r$ respectively. This implies $[\mathbf{F}]_{i,j} = \frac{1}{\sqrt{N_t}} e^{j\psi_{i,j}}$ and $[\mathbf{W}]_{i,j} = \frac{1}{\sqrt{N_r}} e^{j\phi_{i,j}}$ where $\psi_{i,j} \in \mathcal{A}_t$ and $\phi_{i,j} \in \mathcal{A}_r$. Vectorizing \eqref{eq:sys_model} and utilizing the Kronecker product identity $\underline{\mathbf{ABC}} = (\mathbf{C}^{T} \otimes \mathbf{A})\underline{\mathbf{B}}$, we obtain
\begin{equation} \label{eq:kron_sys_model}
    \underline{\mathbf{y}} = (\mathbf{S}^T\mathbf{F}^T \otimes \mathbf{W}^H) \underline{\mathbf{H}} + (\mathbf{I}_{N_p} \otimes \mathbf{W}^H) \underline{\mathbf{n}},
\end{equation}
where $\underline{\mathbf{y}} \in \mathbb{C}^{N_{\mathrm{s}}N_{\mathrm{p}} \times 1}$, $\underline{\mathbf{H}} \in \mathbb{C}^{N_{\mathrm{r}}N_{\mathrm{t}} \times 1}$ and $\underline{\mathbf{n}} \in \mathbb{C}^{N_{\mathrm{r}}N_{\mathrm{p}} \times 1}$. Denote $\mathbf{A} = (\mathbf{S}^T\mathbf{F}^T \otimes \mathbf{W}^H)$. $\mathbf{A}$ has dimensions $N_sN_p \times N_tN_r$, however its rank is bounded by $N_s^2$, since $\mathbf{F}$ can have atmost $N_s$ independent columns. Consequently, if $N_sN_p < N_tN_r$, the system of equations given by \eqref{eq:kron_sys_model} does not have a unique solution, and we cannot perform traditional least-squares (LS) estimation to recover $\underline{\mathbf{H}}$. In other words, recovering $\underline{\mathbf{H}}$ from $\underline{\mathbf{y}}$ given $\mathbf{A}$ is an inverse problem. In Section~\ref{sec:gce}, we will describe an algorithm to solve this inverse problem using a deep generative prior.

\section{Generative Channel Estimation (GCE)}
\label{sec:gce}
Deep generative models $\mathbf{G}$ are feed-forward neural networks (NN) that take as input a low dimensional vector $\mathbf{z} \in \mathbb{R}^d$ and output high dimensional matrices $\mathbf{G}(z) \in \mathbb{R}^{c \times l \times w}$ where $d \ll clw$. Such a model can be trained to take a i.i.d. Gaussian vector $\mathbf{z}$ as input and produce samples from complicated distributions, such as human faces \cite{radford2015unsupervised}. One powerful method for training generative models is using Generative Adversarial Networks (GAN) \cite{goodfellow2014generative}. 

In \cite{balevi2020high}, we developed an algorithm that utilized compressed sensing using deep generative models \cite{bora2017compressed} to perform MIMO channel estimation. We trained $\mathbf{G}$ to output channel realizations $\mathbf{H}$ from a given distribution, and then utilized $\mathbf{G}$ to recover $\mathbf{H}$ from compressive pilot measurements $\underline{\mathbf{y}}$. However, we experimentally demonstrated that a reduced antenna spacing of $\lambda_c/10$ in the antenna arrays at the transmitter and receiver was key to training $\mathbf{G}$ successfully. We attributed this to the high spatial correlation generated in channel realizations by such an antenna spacing, making it easier for $\mathbf{G}$ to learn the underlying channel distribution. 

In this paper, we will utilize the following key insight  to output channel realizations with the conventional $\lambda_c/2$ antenna spacing: \textit{beamspace representation of mmWave MIMO channels have high spatial correlation due to clustering in the angular domain.} To be precise, assuming uniformly spaced linear arrays at the transmitter and receiver, the array response matrices are given by the unitary DFT matrices $\mathbf{A}_{\mathrm{T}} \in \mathbb{C}^{N_{\mathrm{t}} \times N_{\mathrm{t}}}$ and $\mathbf{A}_{\mathrm{R}} \in \mathbb{C}^{N_{\mathrm{r}} \times N_{\mathrm{r}}}$ respectively. Then, we can represent $\mathbf{H}$ as
\begin{equation} \label{eq:virtual_channel_rep}
\begin{aligned}
    \mathbf{H} &= \mathbf{A}_{\mathrm{R}}\mathbf{H}_\mathrm{v}\mathbf{A}_{\mathrm{T}}^H \\
    \underline{\mathbf{H}} &= ((\mathbf{A}_{\mathrm{T}}^H)^T \otimes \mathbf{A}_{\mathrm{R}})\underline{\mathbf{H}_\mathrm{v}}.
\end{aligned}
\end{equation}
We will train $\mathbf{G}$ to output samples of $\mathbf{H}_{\mathrm{v}}$ i.e. $\mathbb{P}_{\mathbf{G}}$ converges to  $\mathbb{P}_{\mathbf{H}_{\mathrm{v}}}$ as GAN training progresses. Note that unlike compressed sensing (CS) algorithms that require $\mathbf{H}_{\mathrm{v}}$ to have a small number of non-zero measurements in order to be recoverable \cite{donoho2005stable}, we impose no such constraints on $\mathbf{G}(\mathbf{z})$. However, we will experimentally validate in Section~\ref{sec:results} that outputting $\mathbf{H}_{\mathrm{v}}$ instead of $\mathbf{H}$ enables $\mathbf{G}$ to directly exploit the clustering in angular domain to successfully learn a generative prior.
\begin{figure}
    \centering
    \includegraphics[width=5in]{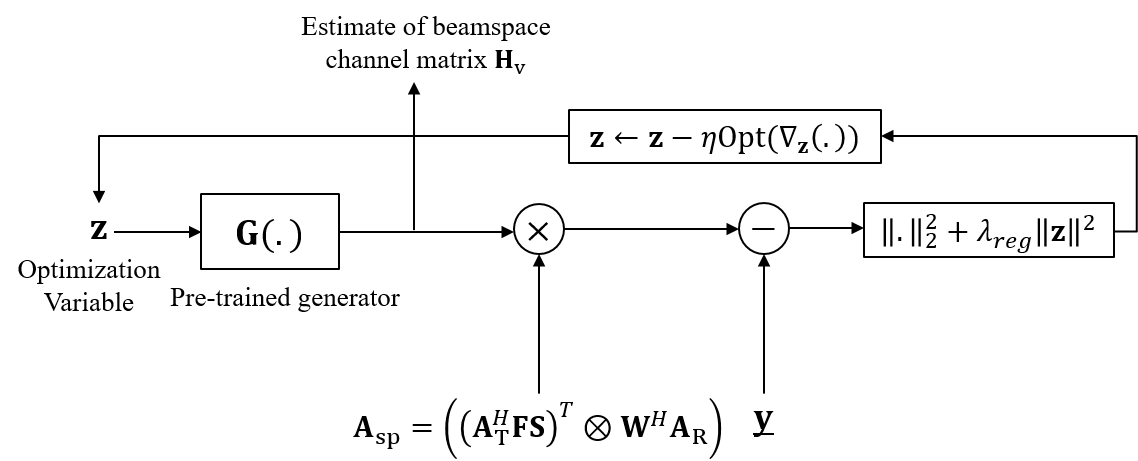}
    \caption{Generative Channel Estimation Framework. $\eta$ denotes the step size and $\mathrm{Opt}$ denotes any first order optimizer.}
    \label{fig:gce}
\end{figure}

Subsequently, given a trained generator $\mathbf{G}$ and pilot measurements $\mathbf{\underline{y}}$ as defined in \eqref{eq:kron_sys_model}, we will utilize GCE to solve the following optimization problem 
\begin{equation} \label{eq:gce}
    \mathbf{z}^* = \underset{\mathbf{z} \in \mathbb{R}^d} {\mathrm{arg\ min\ }} \hspace{0.05 in} ||\underline{\mathbf{y}} - \mathbf{A}_{\mathrm{sp}} \mathbf{\underline{G}}(\mathbf{z})||_{2}^2 + \lambda_{\mathrm{reg}} ||\mathbf{z}||_{2}^2,
\end{equation}
where $\mathbf{A}_{\mathrm{sp}} = (\mathbf{A}_{\mathrm{T}}^H\mathbf{F}\mathbf{S})^T \otimes \mathbf{W}^H\mathbf{A}_{\mathrm{R}}$ and $\lambda_{\mathrm{reg}}$ is a regularization parameter. An illustration of the framework is shown in Fig.~\ref{fig:gce}. The beamspace channel estimate is then given by $\mathbf{H}_\mathrm{v,est} = \mathbf{G}(\mathbf{z}^*)$. The performance metric used to assess the quality of $\mathbf{H}_\mathrm{v,est}$ is the normalized mean square error (NMSE), defined as 
\begin{equation} \label{eq:NMSE}
\text{NMSE} = \mathbb{E}\left[\frac{||\mathbf{H}_\mathrm{v}-\mathbf{H}_\mathrm{v,est}||_2^2}{||\mathbf{H}_\mathrm{v}||_2^2}\right],
\end{equation}
where the expected value is over the underlying probability distribution of the channel. It should be noted that since $\mathbf{A}_{\mathrm{R}}$ and $\mathbf{A}_{\mathrm{T}}$ are unitary matrices, the NMSE is the same in the beamspace $(\mathbf{H}_\mathrm{v,est})$ and spatial $(\mathbf{H}_\mathrm{est})$ domain. The NMSE, as defined in \eqref{eq:NMSE}, will serve as a measure of the quality of the generator $\mathbf{G}$ trained by a GAN throughout the paper. 

\section{GAN architectures} \label{sec:gan_arch}
In this section, we will present three different GAN architectures - (i) Wasserstein GAN with Gradient Penalty (WGAN-GP), (ii) Conditional Wasserstein GAN (CWGAN) and (iii) a new novel GAN architecture called Pilot GAN. We will present all three architectures in the context of narrowband MIMO channel generation (refer Remark \ref{remark:wideband}), in keeping with the GCE framework outlined in Section~\ref{sec:gce}. In other words, we will assume that the generator in all GAN architectures will output the beamspace MIMO channel representation given by \eqref{eq:virtual_channel_rep}. Then we will describe how the second and third GAN architectures can be combined to form Pilot Conditional GAN, that in combination with the LOS Predictor presented in Section~\ref{subsec:pcgan}, can be trained OTA at the receiver across a range of channel models.
\begin{remark} \label{remark:wideband}
\textit{Extension to wideband channels could be performed by using GCE for each pilot subcarrier and subsequently interpolating between subcarriers. Comprehensive practical evaluation in \cite{zhang2020deepwiphy} has shown minimal performance gain in employing DL-based CE techniques over the OFDM resource grid.}
\end{remark}

\subsection{Wasserstein GAN with Gradient Penalty}
A Wasserstein Generative Adversarial Network (WGAN) \cite{arjovsky2017wasserstein} consists of two deep neural networks - a generator $\mathbf{G}(.;\theta_g)$ and a critic\footnote{WGAN does not refer to this as a discriminator since it is not trained to classify i.e. there is no Sigmoid activation in the final layer of a critic NN.} $\mathbf{D}(.;\theta_d)$ - whose weights are optimized so as to solve the following min-max problem:
\begin{equation} \label{eq:wgan}
    \underset{\mathbf{G}}{\mathrm{min}}~\underset{\mathbf{D} \in \mathcal{D}}{\mathrm{max}} ~ \mathbb{E}_{\mathbf{x}\sim \mathbb{P}_{r}(\mathbf{x})} \mathbf{D}(\mathbf{x}) 
    - \mathbb{E}_{z\sim \mathbb{P}_{z}(z)}\mathbf{D}(\mathbf{G}(\mathbf{z})),
\end{equation}
where $\mathcal{D}$ is the set of 1-Lipschitz functions\footnote{A differentiable function $f$ is said to be $\alpha$-Lipschitz if $||\nabla_xf(x)||_2 \leq \alpha$ $\forall$ $x$.} and $\mathbb{P}_{r}$ is the data distribution. The original GAN \cite{goodfellow2014generative} is famously known to suffer from mode collapse \cite{srivastava2017veegan} i.e. $\mathbb{P}_{\mathbf{G}}$ collapses to a delta function centered around the mode of the input data distribution. In \cite{arjovsky2017wasserstein}, they attribute this behaviour to the use of the KL (Kullback-Leibler) or JS (Jensen-Shannon) divergence during training, and instead propose using the Wasserstein-1 distance. They show that sequences of probability distributions that converge under Wasserstein-1 are a superset of sequences convergent under KL or JS divergence, making WGAN more robust to mode collapse.

WGAN with Gradient Penalty (WGAN-GP) \cite{gulrajani2017improved} improves the performance of WGAN by incorporating a penalty on the gradient norm for random samples $\hat{\mathbf{x}} \sim \mathbb{P}_{\hat{\mathbf{x}}}$ as a soft version of the Lipschitz constraint. In this paper, we will be utilizing both WGAN and WGAN-GP, depending on which training procedure yields better performance (further details in Section~\ref{sec:results}). In case of WGAN, the Lipschitz constraint $\mathbf{D} \in \mathcal{D}$ in \eqref{eq:wgan} is enforced by clipping $\theta_d$ to be between $(-\tau,\tau)$, where $\tau$ is the clipping constant. In case of WGAN-GP, the Lipschitz constraint is enforced by adding
\begin{equation}
    L_{\mathrm{GP}}(\theta_d) = \mathbb{E}_{\hat{\mathbf{x}}\sim\mathbb{P}_{\hat{\mathbf{x}}}} \big[ (||\nabla_{\hat{\mathbf{x}}} \mathbf{D}(\hat{\mathbf{x}};\theta_d)||_2 - 1)^2 \big]
\end{equation}
to the objective in \eqref{eq:wgan}, where $\hat{\mathbf{x}}$ are points sampled uniformly along straight lines joining pair of points sampled from the data distribution $\mathbb{P}_r$ and the generator distribution $\mathbb{P}_{\mathbf{G}}$, since enforcing the gradient penalty over all possible inputs to $\mathbf{D}$ is intractable \cite{gulrajani2017improved}.  

A unified algorithm capturing the training of both WGAN and WGAN-GP in the context of channel generation is outlined in Algorithm~\ref{alg:WGAN_GP_training} by utilizing an indicator $\mathbbm{1}_{\mathrm{GP}}$ to indicate if WGAN-GP was chosen or not. We update the critic and generator weights, $\theta_d$ and $\theta_g$, using calls to Subroutine~\ref{subrout:d} and~\ref{subrout:g} respectively, in Algorithm~\ref{alg:WGAN_GP_training}. Both subroutines will be used in all subsequent GAN training algorithms. As part of updating the critic in Algorithm~\ref{alg:WGAN_GP_training}, we note that the gradient penalty is enforced only along straight lines between pairs of points sampled from the beamspace channel distribution $\mathbb{P}_{\mathbf{H}_{\mathrm{v}}}$ and the generator distribution $\mathbb{P}_{\mathbf{G}}$. In Section~\ref{subsec:results_wgan_gp}, we will highlight the NMSE improvement obtained by using WGAN-GP instead of WGAN.
\begin{algorithm} 
\DontPrintSemicolon
\SetAlgoHangIndent{0pt}
\setstretch{1}
\For{number of training iterations}{
 \For{$n_{\mathrm{d}}$ iterations}{
    Sample minibatch of $m$ beamspace channel realizations $\{\mathbf{H}_{\mathrm{v}}^{(i)}\}_{i=1}^{m} \sim \mathbb{P}_{\mathbf{H}_{\mathrm{v}}}$, latent variables $\{\mathbf{z}^{(i)}\}_{i=1}^{m} \sim \mathbb{P}_z$ and random numbers $\{\epsilon^{(i)}\}_{i=1}^{m} \sim U[0,1]$.\;
    $\tilde{\mathbf{H}}_\mathrm{v} = \mathbf{G}(\mathbf{z};\theta_g)$.\;
    $\hat{\mathbf{H}}_\mathrm{v} = \epsilon  \mathbf{H}_{\mathrm{v}} + (1-\epsilon)  \tilde{\mathbf{H}}_\mathrm{v}$.\;
    $\theta_d = $ \texttt{Update\_D}$(\tilde{\mathbf{H}}_\mathrm{v},\mathbf{H}_\mathrm{v},\hat{\mathbf{H}}_\mathrm{v},\mathbbm{1}_{\mathrm{GP}},m,\gamma,\tau,\beta;\theta_d)$\;
    }
    Sample minibatch of $m$ latent variables $\{\mathbf{z}^{(i)}\}_{i=1}^{m} \sim \mathbb{P}_z$.\;
    $\theta_g = $ \texttt{Update\_G}$(\mathbf{G(z;\theta_g)},m,\gamma;\theta_g)$
    }
\caption[caption]{Wasserstein GAN} \label{alg:WGAN_GP_training}
\end{algorithm}
\begin{subroutine} 
\DontPrintSemicolon
\SetAlgoHangIndent{0pt}
\begin{fleqn}[\parindent]
\begin{equation*}
\begin{split}
&L(\theta_d) = \frac{1}{m} \sum_{i=1}^{m} \mathbf{D}(\mathbf{x}^{(i)}_{\mathbf{G}};\theta_d) - \mathbf{D}(\mathbf{x}^{(i)}_r;\theta_d) + \beta \mathbbm{1}_{\mathrm{GP}} ( || \nabla_{\mathbf{x}^{(i)}_{r\mathbf{G}}} D( \mathbf{x}^{(i)}_{r\mathbf{G}};\theta_d) ||_2 - 1)^2 \\
&\theta_d = \theta_d - \gamma \mathrm{RMSProp}(\nabla_{\theta_d} L(\theta_d)) \\
&\theta_d = \mathbbm{1}_{\mathrm{GP}} \theta_d + (1 - \mathbbm{1}_{\mathrm{GP}}) \mathrm{clip}(\theta_d,-\tau,\tau)
\end{split}
\end{equation*}
\end{fleqn}
\caption{$\theta_d = $ \texttt{Update\_D}$(\mathbf{x}_{\mathbf{G}},\mathbf{x}_r,\mathbf{x}_{r\mathbf{G}},\mathbbm{1}_{\mathrm{GP}},m,\gamma,\tau,\beta;\theta_d)$} \label{subrout:d}
\end{subroutine}
\begin{subroutine}
\DontPrintSemicolon
\SetAlgoHangIndent{0pt}
\begin{fleqn}[\parindent]
\begin{equation*}
\begin{split}
&L(\theta_g) = \frac{1}{m} \sum_{i=1}^{m} -\mathbf{D}(\mathbf{x}^{(i)}_{\mathbf{G}})  ~~~~~~~~\text{\footnotesize{$\%$ \texttt{$\mathbf{x}_{\mathbf{G}}$ will be a function of $\theta_g$}}}\\
&\theta_g = \theta_g - \gamma \mathrm{RMSProp}(\nabla_{\theta_g} L(\theta_g)) \\
\end{split}
\end{equation*}
\end{fleqn} 
\caption{$\theta_g = $ \texttt{Update\_G}$(\mathbf{x}_{\mathbf{G}},m,\gamma;\theta_g)$} \label{subrout:g}
\end{subroutine}

\subsection{Conditional Wasserstein GAN} \label{subsec:cwgan_gp}
The WGAN in \cite{balevi2020high} was trained on channel realizations drawn from a single distribution which was characterized by a very strong Line-of-Sight (LOS) component. Moreover, training on such a channel distribution provides no indication of the generator's ability to learn more complex multi-path channels. In this section, we will present a Conditional WGAN (CWGAN) design that will have the ability to be trained on channel realizations drawn from a plurality of distributions, each yielding channel realizations with varying degrees of approximate sparsity in the beamspace domain.

\begin{figure}
    \centering
    \includegraphics[width=5in]{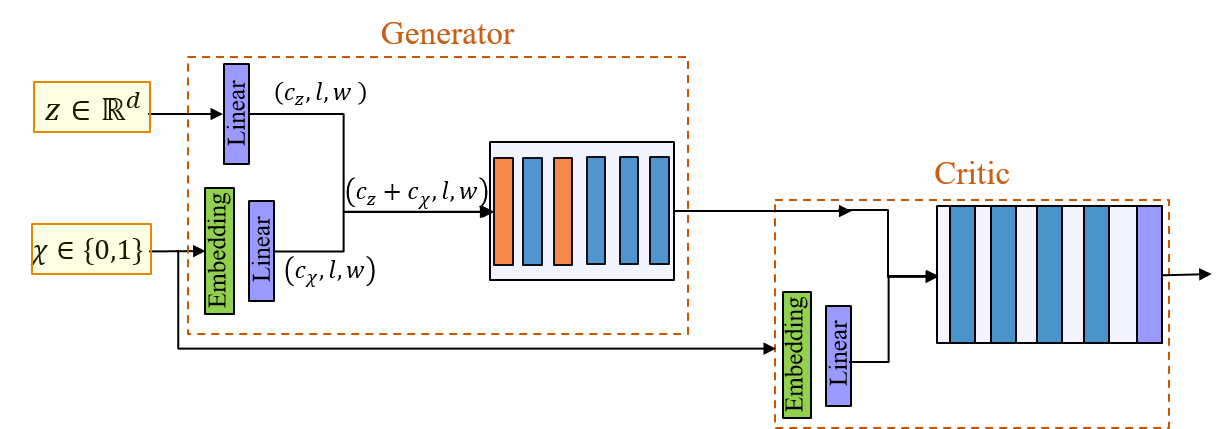}
    \caption{Architecture of CWGAN to highlight how the condition $\chi$ is input to both $\mathbf{G}$ and $\mathbf{D}$. Only the path of the generated channel realizations $\mathbf{G}(z,\chi)$ through the GAN is shown here, a similar path can be traced for $\{\mathbf{H}_{v}^{(i)}\}_{i=1}^{m}$.}
    \label{fig:cgan}
\end{figure}

We combine the architecture of Conditional GAN \cite{mirza2014conditional} with the training procedure of WGAN outlined in Algorithm~\ref{alg:WGAN_GP_training} to develop CWGAN. While we will describe the NN architectures of the generator and discriminator in Section~\ref{subsec:nn_arch}, we have depicted how they are modified to accept as an input a condition in Fig.~\ref{fig:cgan}. For now, we assume that we are provided with a binary label $\chi$ indicating whether the channel we are trying to estimate is LOS ($\chi=1$) or NLOS ($\chi=0$). We will present different techniques to overcome this assumption during training and testing in Section~\ref{subsec:pcgan}. The condition $\chi$ is then passed through a learnable Embedding\footnote{https://pytorch.org/docs/stable/generated/torch.nn.Embedding.html} layer, that embeds an integer as a high dimensional vector, followed by a linear and reshaping layer that has output dimensions $(c_{\chi},l,w)$ as shown in Fig.~\ref{fig:cgan}. This embedded output is then appended along the channel dimension to $\mathrm{Linear}(\mathbf{z})$ of shape $(c_z,l,w)$ to yield an input of size $(c_z+c_{\chi},l,w)$ that is passed through the remaining deep convolutional generative network. A similar procedure is followed while inputting $\chi$ to the critic.

We now need to modify the WGAN training procedure outlined in Algorithm~\ref{alg:WGAN_GP_training} to incorporate the conditional input. To this end, we simply sample $\{\mathbf{H}_{\mathrm{v}}^{(i)},\chi^{(i)}\}_{i=1}^{m} \sim \mathbb{P}_{\mathbf{H}_{\mathrm{v}}}$ and utilize $\{\chi^{(i)}\}_{i=1}^{m}$ as input to $\mathbf{G}$ for computing $\tilde{\mathbf{H}}_{\mathrm{v}}$ and as input to $\mathbf{D}$ for computing $L(\theta_d)$. Despite $\{\chi^{(i)}\}_{i=1}^{m}$ being input to $\mathbf{D}$, the derivative $\nabla_{\hat{\mathbf{H}}^{(i)}_\mathrm{v}}$ continues to remain only w.r.t $\hat{\mathbf{H}}^{(i)}_\mathrm{v}$. Subsequently, we randomly sample $m$ labels in $\{0,1\}$ from a Bernoulli $\mathrm{Ber}(0.5)$ distribution and utilize these as the conditional input to both $\mathbf{G}$ and $\mathbf{D}$ for computing $L(\theta_g)$.

\subsection{Pilot GAN} \label{subsec:pilot_gan}
A large dataset of clean channel realizations was required to train a WGAN in \cite{balevi2020high}. Such a dataset can only be generated offline via simulation tools such as the MATLAB 5G Toolbox or Wireless InSite\footnote{https://www.remcom.com/wireless-insite-em-propagation-software/} and limits the applicability and adaptability of the trained generative models in practical deployments. Instead, we need to develop a  framework to train GANs directly from noisy pilot measurements $\mathbf{\underline{y}}$, as defined in \eqref{eq:kron_sys_model}. To this end, we will combine LS CE (refer Remark \ref{remark:ls_ambgan}) and Ambient GAN \cite{bora2018ambientgan} in order to train a GAN from full rank received pilot measurements and call this framework Pilot GAN. First we present a brief overview of Ambient GAN and the conditions under which it can provably recover the true underlying data distribution.

\vspace{2mm}
\noindent
\textbf{Ambient GAN}: Ambient GAN \cite{bora2018ambientgan} was designed with the objective of training a GAN given only lossy measurements of images from the distribution of interest. It was based on the following key idea: rather than distinguish a real image from a generated image as is traditionally done in a GAN, the Ambient GAN discriminator must distinguish a real measurement from a simulated measurement of a generated image, assuming the measurement process is known. They considered a wide variety of measurement processes, for example convolution with a kernel, additive noise, and projection on a random Gaussian vector, and demonstrated that Ambient GAN trained a generator that produced images with good visual quality in spite of the noisy measurement processes.

The space of permissible measurement processes was determined by the following condition: there should be a \textit{unique} data distribution $\mathbb{P}_r$ consistent with the observed measurement distribution $\mathbb{P}_y$. In other words, there must be an invertible mapping between $\mathbb{P}_r$ and $\mathbb{P}_y$ even though the map from an individual image to its measurement may not be invertible.

\begin{figure} 
    \centering
    \includegraphics[width=5in]{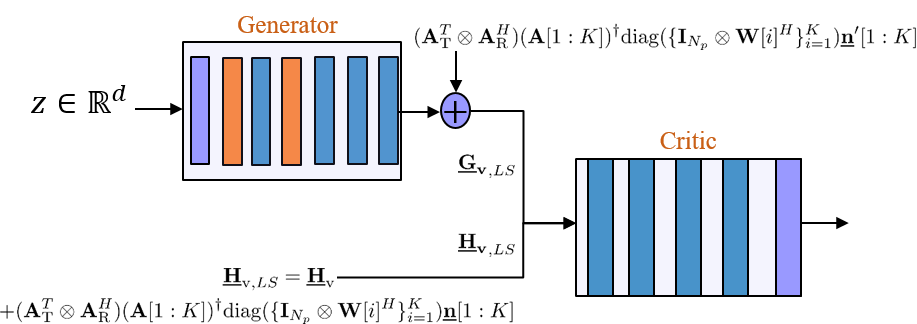}
    \caption{Architecture of Pilot GAN. Correlated complex noise is added to the output of $\mathbf{G}$ before passing it to the critic. Complex observations are stacked as real and imaginary along the channel dimension before being input to the critic.}
    \label{fig:pilot_gan}
\end{figure}

\vspace{2mm}
\noindent
\textbf{Combining LS CE and Ambient GAN}: In Section~\ref{sec:sys_model}, we described how the rank of the measurement matrix $\mathbf{A}$ is bounded by $N_s^2$, even if $N_p > N_s$. Hence, let us assume that $N_p = N_s$. In order to obtain a set of full rank measurements, we will consider $K = \lceil N_t/N_p \rceil$ consecutive transmissions each with a different $(\mathbf{F}[i],\mathbf{S}[i],\mathbf{W}[i])$ triplet. Assuming the channel $\mathbf{H}$ remains constant over $K$ consecutive transmissions, we can stack the $K$ received signals to write $\mathbf{\underline{y}}[1:K]$ as
\begin{equation} \label{eq:pilot_gan_stack_sys_model}
    \begin{bmatrix}
    \mathbf{S}[1]^T\mathbf{F}[1]^T \otimes \mathbf{W}[1]^H \\
    \vdots \\
    \mathbf{S}[K]^T\mathbf{F}[K]^T \otimes \mathbf{W}[K]^H \end{bmatrix} \mathbf{\underline{H}} + \begin{bmatrix}
    \mathbf{I}_{N_p} \otimes \mathbf{W}[1]^H & \mathbf{0} & \cdots & \mathbf{0} \\
    \vdots  & \vdots  & \ddots & \vdots  \\
    \mathbf{0} & \mathbf{0} & \cdots & \mathbf{I}_{N_p} \otimes \mathbf{W}[K]^H 
    \end{bmatrix} \begin{bmatrix}
    \mathbf{\underline{n}}[1] \\
    \vdots \\
    \mathbf{\underline{n}}[K] \end{bmatrix} 
\end{equation}
By appropriately sampling the entries of $\mathbf{F}[i]$, $\mathbf{S}[i]$ and $\mathbf{W}[i]$ for all $i \in [K]$, we can ensure that the stacked measurement matrix $\mathbf{A}[1:K]$ has full rank $N_tN_r$. Hence, the beamspace LS channel estimate is given by
\begin{align} 
    \label{eq:full_rank_ls_est}
    \mathbf{\underline{H}}_{\mathrm{v},LS} &= \mathbf{\underline{H}}_\mathrm{v} + \boldsymbol{\underline{\zeta}},
\end{align}
where $\boldsymbol{\underline{\zeta}}$ is a zero-mean complex Gaussian random vector of covariance $\mathbf{\Sigma}$ such that 
\begin{equation} \label{eq:h_ls}
    \mathbf{\Sigma}^{\frac{1}{2}} =  (\mathbf{A}^T_{\mathrm{T}} \otimes \mathbf{A}^H_{\mathrm{R}}) (\mathbf{A}[1:K])^{\dagger} \mathrm{diag}(\{\mathbf{I}_{N_p} \otimes \mathbf{W}[i]^H\}_{i=1}^{K}), 
\end{equation}
where $(\mathbf{A}[1:K])^{\dagger}$ denotes the pseudo-inverse of $\mathbf{A}[1:K]$. The pseudo-inverse provides for $K > N_t/N_p$. If $K = N_t/N_p$, this simply reduces to $(\mathbf{A}[1:K])^{-1}$. In Ambient GAN terms, $\mathbf{\underline{H}}_{\mathrm{v},LS}$ is the lossy measurement, while the measurement process is additive complex Gaussian noise with covariance $\mathbf{\Sigma}$. Subsequently, training an Ambient GAN now simply involves adding noise of the same covariance $\mathbf{\Sigma}$, as given by \eqref{eq:h_ls}, to $\mathbf{\underline{G}}(\mathbf{z})$. Given that these are pilot transmissions, we can assume that the transmitter and receiver have agreed upon a set of $K$ $(\mathbf{F}[i],\mathbf{S}[i],\mathbf{W}[i])$ triplets, that are known to both. Since $\mathbf{\Sigma} = f\big(\{(\mathbf{F}[i],\mathbf{S}[i],\mathbf{W}[i])\}_{i=1}^{K}\big)$, we can easily generate complex Gaussian noise of the desired covariance $\mathbf{\Sigma}$ and add it to the output of the generator to generate fake beamspace LS channel estimates $\mathbf{\underline{G}}_{\mathrm{v},LS}$ as shown in Fig.~\ref{fig:pilot_gan}. Having thus performed LS CE to obtain the dataset $\{\mathbf{\underline{H}}_{\mathrm{v},LS}^{(i)}\}$, and added correlated noise to the generator output to obtain $\{\mathbf{\underline{G}}_{\mathrm{v},LS}^{(i)}\}$, we can perform the same WGAN training procedure outlined in Algorithm~\ref{alg:WGAN_GP_training} to train the Pilot GAN.

Throughout the exposition of Pilot GAN, we have assumed full-rank pilot measurements in order to perform LS CE. This can be attributed to the usage of Ambient GAN which requires there to be a unique distribution $\mathbb{P}_{\mathbf{H}_{\mathrm{v}}}$ consistent with the observed pilot measurements $\mathbb{P}_{\mathbf{Y}}$. It can be easily shown that additive noise preserves distributional invertibility (refer Section 10.3 in \cite{bora2018ambientgan}). However, if we were to multiply $\mathbf{H}_{\mathrm{v}}$ by a compressive measurement matrix $\mathbf{A}$, we cannot express $\mathbb{P}_{\mathbf{Y}}$ as a unique function of $\mathbb{P}_{\mathbf{H}_{\mathrm{v}}}$, even in the noiseless setting (refer Appendix~\ref{subsec:proof}). Hence, we have utilized full-rank pilot measurements for \textit{training} Pilot GAN, even though it is important to note that during \textit{inference} the generative model uses compressive measurements for channel estimation.

\vspace{2mm}
\noindent
\textbf{Federated Pilot GAN:} In a setting where $U > 1$ UEs are connected to a single BS, we can easily extend the Pilot GAN training to a federated implementation by utilizing the following  insight -- only the critic training requires the $\{\mathbf{\underline{H}}_{\mathrm{v},LS}^{(i)}\}$ dataset, the generator training on the other hand does not \cite{augenstein2019generative}. Hence we can train the generator $\mathbf{G}$ at the BS in a centralized fashion, while the critic training is carried out locally at the UEs and its gradient updates are averaged at the BS. The training procedure for Federated Pilot GAN is presented in Algorithm~\ref{alg:FedGAN_training}. In summary, we assume that each UE $u$ periodically accumulates a small dataset of $D$ beamspace LS channel estimates and uses that to train its critic $\mathbf{D}_u$. After each UE has performed $n_d$ critic training iterations, it transmits the critic weights $\theta_{d,u}$ to the BS where they are averaged to obtain $\theta_d$ and used to train $\mathbf{G}$, following which both $\theta_g$ and $\theta_d$ are broadcast to all UEs. Since $\mathbf{G}$ is typically significantly more complex than $\mathbf{D}$ (refer Table~\ref{tab:nn_size}), this ensures that the bulk of the training is carried out at the BS.

It should be noted that we assume error free transmission of the federated updates in both the uplink and downlink, which can be achieved by utilizing sub-6 GHz channels for these transmissions as is commonly done in Carrier Aggregation in 5G systems, i.e. utilizing a lower bandwidth channel at a lower $f_c$ along with mmWave.
\begin{algorithm}
\DontPrintSemicolon
\SetAlgoHangIndent{0pt}
\setstretch{1}
\For{number of rounds}{
    UE $u$ accumulates $D$ beamspace LS channel estimates $\{\mathbf{H}_{\mathrm{v},LS,u}^{(i)}\}_{i=1}^{D}$ $\forall u \in [U]$\;
    \For{$l$ training iterations}{
        BS broadcasts $\theta_g$ and $\theta_d$ to all UEs\;
        \For{$u = 1:U$}{
            $\theta_{d,u} = \theta_d$\;
            \For{$n_d$ iterations}{
                Sample minibatch $\mathbf{B}_u$ of size $m/U$ from $\{\mathbf{H}_{\mathrm{v},LS,u}^{(i)}\}_{i=1}^{D}$, latent variables $\{\mathbf{z}^{(i)}\}_{i=1}^{m/U} \sim \mathbb{P}_z$ and random numbers $\{\epsilon^{(i)}\}_{i=1}^{m/U} \sim U[0,1]$.\;
                $\mathbf{\underline{G}}_{\mathrm{v},LS} = \mathbf{\underline{G}}(\mathbf{z}) + \mathcal{CN}(\mathbf{0},\mathbf{\Sigma})$, with $\mathbf{\Sigma}$ given by \eqref{eq:h_ls}. \;
                $\hat{\mathbf{B}}_{u} = \epsilon \mathbf{B}_u + (1-\epsilon) \mathbf{G}_{\mathrm{v},LS}$.\;
                $\theta_{d,u} =$ \texttt{Update\_D}$(\mathbf{\underline{G}}_{\mathrm{v},LS},\mathbf{B}_u,\hat{\mathbf{B}}_{u},\mathbbm{1}_{\mathrm{GP}},m/U,\gamma,\tau,\beta;\theta_{d,u})$
            }
            UE $u$ sends $\theta_{d,u}$ in uplink to BS\;
        }
        $\theta_d = (1/U)\sum_{u=1}^{U} \theta_{d,u}$\;
        Sample minibatch of $m$ latent variables $\{\mathbf{z}^{(i)}\}_{i=1}^{m} \sim \mathbb{P}_z$\;
        $\mathbf{\underline{G}}_{\mathrm{v},LS} = \mathbf{\underline{G}}(\mathbf{z};\theta_g) + \mathcal{CN}(\mathbf{0},\mathbf{\Sigma})$, with $\mathbf{\Sigma}$  given by \eqref{eq:h_ls}.\;
        $\theta_g =$ \texttt{Update\_G}$\big(\mathbf{G}_{\mathrm{v},LS},m,\gamma;\theta_g\big)$
    }
}
\caption[caption]{Federated Pilot GAN with $U$ UEs.} \label{alg:FedGAN_training}
\end{algorithm}

\begin{remark} \label{remark:ls_ambgan}
\textit{Unlike channel estimation, generative model training is not a latency sensitive operation. Consequently, waiting $K$ transmissions to obtain a LS channel estimate to train Pilot GAN should not be misinterpreted as implying that LS channel estimation can be used to benchmark the performance of GCE. }
\end{remark}

\subsection{Pilot Conditional GAN with LOS Predictor} \label{subsec:pcgan}
Having presented a framework for training a generative model from pilot measurements, the sensible next step would be to combine the CWGAN framework presented in Section~\ref{subsec:cwgan_gp} with the Pilot GAN framework of Section~\ref{subsec:pilot_gan} to train a generative model from pilot measurements of all LOS and NLOS channels. However, prior to doing that, we will present a simple supervised learning based technique to determine the LOS/NLOS label $\chi$ from the received full-rank pilot measurements $\mathbf{\underline{y}}[1:K]$.  

\subsubsection{LOS Predictor} 
Intuitively, we expect to easily distinguish between LOS and NLOS channel realizations by looking at the number and magnitude of non-zero entries in the beamspace representation $\mathbf{H}_{\mathrm{v}}$.
Since we assume full-rank pilot measurements, the beamspace LS channel estimates $\mathbf{H}_{\mathrm{v},LS}$ are noisy versions of $\mathbf{H}_{\mathrm{v}}$. Hence, we simply 
train a NN $\mathcal{L}(.;\theta_L)$ in a supervised fashion to take as input $\mathbf{H}_{\mathrm{v},LS}$ and output $\mathbb{P}(\mathbf{H}_{\mathrm{v},LS} \in \mathrm{LOS})$. Given LOS ground truth labels, such a NN can be trained using the binary cross entropy (BCE) loss\footnote{https://pytorch.org/docs/stable/generated/torch.nn.BCELoss.html} and the condition $\chi$ is given by
\begin{equation} \label{eq:cond_LOS}
    \chi = \frac{1}{2}\big(1+\mathrm{sgn} \big(2 \mathcal{L}(\mathbf{H}_{\mathrm{v},LS};\theta_L) - 1\big)\big).
\end{equation}
It may appear that the supervised training of an LOS Predictor is at odds with the overarching goal of this paper to design an OTA training procedure. However, the minimal loss in accuracy of the LOS predictor during inference, when trained on only a subset of the CDL channel models as demonstrated in Section~\ref{subsec:results_pcgan}, will reinforce the practicality of the proposed approach.

As explained in Section~\ref{subsec:pilot_gan}, we require full rank pilot measurements to perform LS CE. While such pilot measurements would be available during Pilot GAN training, the measurements received while performing GCE would be compressive. However, the requirement of an LOS predictor during GCE can be easily circumvented by modifying the optimization in \eqref{eq:gce} to
\begin{equation} \label{eq:gce_cond}
    \mathbf{z}^* = \underset{\chi \in \{0,1\}} {\mathrm{min\ }} \underset{\mathbf{z} \in \mathbb{R}^d} {\mathrm{arg\ min\ }} \hspace{0.05 in} ||\underline{\mathbf{y}} - \mathbf{A}_{\mathrm{sp}} \mathbf{\underline{G}}(\mathbf{z},\chi)||_{2}^2 + \lambda_{\mathrm{reg}} ||\mathbf{z}||_{2}^2.
\end{equation}
Note that since $\chi \in \{0,1\}$, it only increases the complexity by a constant factor of 2.

\subsubsection{Pilot Conditional GAN}
Having designed a LOS predictor, we now present a unified algorithm to train a generative model from noisy full-rank pilot measurements that is applicable across all MIMO channel models -- PCGAN. The technique is summarized in Algorithm~\ref{alg:PCGAN_training}. 

It is worth emphasising the completely unsupervised and OTA nature of this GAN implementation (aside from the usage of the LOS Predictor). Having received pilot measurements $\mathbf{\underline{y}}[1:K]$, we first compute the beamspace LS channel estimate $\mathbf{H}_{\mathrm{v},LS}$. These channel estimates are used to train a conditional Wasserstein GAN. Subsequently, for CE, we will extract the conditional generator $\mathbf{G}$ and compute the beamspace channel estimate $\mathbf{G}(\mathbf{z}^*)$ from compressive pilot measurements $\mathbf{\underline{y}}$ by solving \eqref{eq:gce_cond}. Hence we have utilized pilot measurements to \textit{train a channel estimator} as well as \textit{perform channel estimation}.

\begin{algorithm} 
\SetAlgoHangIndent{0pt}
\DontPrintSemicolon
\For{number of training iterations}{
 \For{$n_{\mathrm{d}}$ iterations}{
    Sample minibatch of $m$ beamspace LS channel estimates $\{\mathbf{H}_{\mathrm{v},LS}^{(i)}\}_{i=1}^{m}$, latent variables $\{\mathbf{z}^{(i)}\}_{i=1}^{m} \sim \mathbb{P}_z$ and random numbers $\{\epsilon^{(i)}\}_{i=1}^{m} \sim U[0,1]$.\;
    Compute $\{\chi^{(i)}\}_{i=1}^{m}$ using \eqref{eq:cond_LOS}.\;
    $\mathbf{\underline{G}}_{\mathrm{v},LS} = \mathbf{\underline{G}}(\mathbf{z},\chi) + \mathcal{CN}(\mathbf{0},\mathbf{\Sigma})$, with $\mathbf{\Sigma}$ given by \eqref{eq:h_ls}. \;
    $\hat{\mathbf{H}}_{\mathrm{v},LS} = \epsilon  \mathbf{H}_{\mathrm{v},LS} + (1-\epsilon)  \mathbf{G}_{\mathrm{v},LS}$.\;
    $\theta_d =$ \texttt{Update\_D}$\big(\{\mathbf{G}_{\mathrm{v},LS},\chi\},\{\mathbf{H}_{\mathrm{v},LS},\chi\},\{\hat{\mathbf{H}}_{\mathrm{v},LS},\chi\},\mathbbm{1}_{\mathrm{GP}}, m,\gamma,\tau,\beta;\theta_d\big)$\;
    }
    Sample minibatch of $m$ latent variables $\{\mathbf{z}^{(i)}\}_{i=1}^{m} \sim \mathbb{P}_z$ and conditions $\{\chi^{(i)}\}_{i=1}^{m} \sim \mathrm{Ber}(0.5)$. \;
    $\mathbf{\underline{G}}_{\mathrm{v},LS} = \mathbf{\underline{G}}(\mathbf{z},\chi;\theta_g) + \mathcal{CN}(\mathbf{0},\mathbf{\Sigma})$, with $\mathbf{\Sigma}$  given by \eqref{eq:h_ls}.\;
    $\theta_g =$ \texttt{Update\_G}$\big(\{\mathbf{G}_{\mathrm{v},LS},\chi\},m,\gamma;\theta_g\big)$ }
\caption[caption]{Pilot Conditional GAN with LOS Predictor $\mathcal{L}(.;\theta_L)$.} \label{alg:PCGAN_training}
\end{algorithm}

\section{Simulation Details} \label{sec:simulation}

\subsection{Data Generation \& Pre-processing}
MIMO channel realizations have been generated using the MATLAB 5G Toolbox. Specifically, in accordance with 3GPP TR 38.901 \cite{3gpp.38.901}, we have generated training and test datasets consisting of an equal number of realizations of all categories of CDL channels i.e. CDL-A, B, C (which are NLOS) and CDL-D, E (which are LOS). The channel simulation parameters are summarized in Table~\ref{tab:channel_param}. Since we assume a narrowband block fading model in this paper, we simply extract the $(N_r,N_t)$ matrix corresponding to the first subcarrier and first OFDM symbol from the $(14,12,N_r,N_t)$ channel realizations generated by MATLAB. The aforementioned channel realizations will be used to generate LS channel estimates, as defined in \eqref{eq:full_rank_ls_est}, which in turn will be used to populate the datasets employed for simulating the OTA training procedures of Algorithm~\ref{alg:FedGAN_training} and~\ref{alg:PCGAN_training}.
\begin{table}
	\caption[Simulation Parameters] {Data Generation Parameters}
	\label{tab:channel_param}
	\centering
	\begin{tabular}{ |p{4cm}|p{3.1cm}|}
		\hline
		Delay Profile & CDL - A,B,C,D,E \\\hline
		Dataset Size & Train - $6000 \times 5$\\& Test - $50 \times 5$ \\\hline
		Subcarrier Spacing & 15 kHz\\\hline
		$N_{\mathrm{t}}$ & 64 \\\hline
		$N_{\mathrm{r}}$ & 16 \\\hline
	\end{tabular}
	\hspace{0.2in}
	\begin{tabular}{ |p{4cm}|p{2cm}|}
	    \hline
	    Antenna Array Type & ULA\\\hline
 		Antenna Spacing & $\lambda_c/2$\\\hline
		Sampling Rate & 15.36 MHz\\\hline
		Carrier Frequency & 40 GHz\\\hline
		Delay Spread & 30 ns \\\hline
		Doppler Shift & 5 Hz\\\hline
	\end{tabular}
\end{table}

The generator output $\mathbf{G}(z)$ and discriminator input are of size $(2,N_t,N_r)$, where the first dimension allows us to stack the real and imaginary parts. Based on empirical evidence that a GAN is unable to learn mean-shifted distributions \cite{srivastava2017veegan} as well as our own prior experience in training GANs for spatial channel matrix generation \cite{balevi2020high}, it is important to normalize the data used to train a GAN. Given a beamspace channel realization $\mathbf{H}_{\mathrm{v}}$, $\mu[i,j] = \mathbb{E}\big[\mathbf{H}_{\mathrm{v}}[i,j]\big]$, $\mathrm{Re}(\sigma[i,j]) = \big(\mathbb{E}\big[(\mathrm{Re}(\mathbf{H}_{\mathrm{v}}[i,j] - \mu[i,j]))^2\big]\big)^{0.5}$ and $\mathrm{Im}(\sigma[i,j]) = \big(\mathbb{E}\big[(\mathrm{Im}(\mathbf{H}_{\mathrm{v}}[i,j] - \mu[i,j]))^2\big]\big)^{0.5}$, we normalize the matrix element-wise as
\begin{equation} \label{eq:normalize}
    \mathrm{Re}(\mathbf{H}_{\mathrm{v}}[i,j]) \leftarrow \frac{\mathrm{Re}(\mathbf{H}_{\mathrm{v}}[i,j] - \mu[i,j])}{\mathrm{Re}(\sigma[i,j])} \; \; \; \mathrm{Im}(\mathbf{H}_{\mathrm{v}}[i,j]) \leftarrow \frac{\mathrm{Im}(\mathbf{H}_{\mathrm{v}}[i,j] - \mu[i,j])}{\mathrm{Im}(\sigma[i,j])}.
\end{equation}
We will compute a single $\mu[i,j]$ and $\sigma[i,j]$ for each $(i,j) \in [N_t]\times[N_r]$ over the entire training dataset of each experiment. Such statistics can be computed from the dataset of beamspace LS channel estimates $\{\mathbf{H}_{\mathrm{v},LS}^{(i)}\}$ in an OTA training procedure. In lieu of \eqref{eq:normalize}, the operations $\mathbf{G}(.)$ and $\mathbf{D}(.,\chi)$ will implicitly be used to denote $\mathrm{SN}^{-1}(\mathbf{G}(.))$ and $\mathbf{D}(\mathrm{SN}(.),\chi)$ respectively throughout the paper without exception. Here $\mathrm{SN}$ denotes the operation of \textit{stacking} the real and imaginary part followed by \textit{normalization} using $\{\mu[i,j],\sigma[i,j]\}_{i,j}$ and $\mathrm{SN}^{-1}$ corresponds to \textit{unnormalization} followed by \textit{unstacking} to generate a complex-valued output.

\subsection{Neural Network Architectures \& Training Hyperparameters} \label{subsec:nn_arch}
The basic neural network architectures of the generator $\mathbf{G}$, critic $\mathbf{D}$ and LOS Predictor $\mathcal{L}$
have been detailed in Table~\ref{tab:nn_architectures}. All layers and their corresponding parameters are described in notation standard to PyTorch \cite{NEURIPS2019_9015}. Since the critic and the LOS Predictor have similar architectures, we have merged their representations and used the indicator $\mathbbm{1}_{\mathcal{L}}$ to indicate layers that are only in the LOS Predictor and not the critic. In particular, the critic in a WGAN-GP implementation cannot have batch normalization since it invalidates the gradient penalty computation during critic training (refer Section 4 in \cite{gulrajani2017improved}). All BatchNorm2D layers have $\mathrm{momentum}=0.8$ \cite{arjovsky2017wasserstein} and Conv2D layers have $\mathrm{bias} = \mathrm{False}$. We utilize $d = 65$ for $\mathbf{z} \in \mathbb{R}^d$ (refer Appendix~\ref{subsec:optimal_d} for an empirical justification).
\begin{table}
	\caption[NN Parameters] {Architecture Details of Generator $\mathbf{G}$, Critic $\mathbf{D}$ and LOS Predictor $\mathcal{L}$. The layer index is denoted by \textbf{L} in the top left corner.}
	\label{tab:nn_architectures}
	\centering
    \begin{tabular}{|c|c|c|}
        \hline
        \textbf{L} & \textbf{Generator} & \textbf{Critic / LOS Predictor} \\\hline
        \multirow{2}{*}{1} & Linear($65,8N_tN_r$) & Conv2D($2,16,(3,3),\mathrm{stride}=2$)\\\cline{2-3}
        & ReLU & LeakyReLU(0.2) $+$ Dropout($0.25$)\\\hline
        \multirow{1}{*}{2} & Reshape($128,N_t/4,N_r/4$) & Conv2D($16,32,(3,3),\mathrm{stride}=2$)\\\hline
        \multirow{4}{*}{3} & Upsample($2$) & ZeroPad2D($(0,1,0,1)$)\\\cline{2-3}
        & Conv2D($128,128,(4,4)$) & LeakyReLU(0.2)$+$Dropout($0.25$)$+\mathbbm{1}_{\mathcal{L}}$BatchNorm2D($32$)\\\cline{2-3}
        & BatchNorm2D($128$) & Conv2D($32,64,(3,3),\mathrm{stride}=2$)\\\cline{2-3}
        & ReLU & LeakyReLU(0.2)$+$Dropout($0.25$)$+\mathbbm{1}_{\mathcal{L}}$BatchNorm2D($64$)\\\hline
        \multirow{4}{*}{4} & Upsample($2$) & Conv2D($64,128,(3,3)$)\\\cline{2-3}
        & Conv2D($128,128,(4,4)$) & LeakyReLU(0.2)$+$Dropout($0.25$)$+\mathbbm{1}_{\mathcal{L}}$BatchNorm2D($128$)\\\cline{2-3}
        & BatchNorm2D($128$) & Flatten\\\cline{2-3}
        & ReLU & Linear($3456,1$)\\\hline
        \multirow{1}{*}{5} & Conv2D($128,2,(4,4)$) & $\mathbbm{1}_{\mathcal{L}}$ Sigmoid\\\hline
    \end{tabular}
\end{table}

In order to extend the generator and critic architectures to the conditional setting, we employ an Embedding$(2,10)$ layer in both. This layer learns a $10$-dimensional embedding for $\chi=0$ and $\chi=1$. Subsequently, $\mathbf{G}$ passes this embedding through Linear$(10,N_tN_r/16)$ and Reshape$(1,N_t/4,N_r/4)$ before concatenating it to Linear$(\mathbf{z})$ of size $(127,N_t/4,N_r/4)$ as shown in Fig.~\ref{fig:cgan}. Similarly, $\mathbf{D}$ passes the embedding through Linear$(10,N_tN_r)$ and Reshape$(1,N_t,N_r)$ before concatenating it to the input of size $(2,N_t,N_r)$. Table~\ref{tab:nn_size} contains the total number of parameters in $\mathbf{G}$ and $\mathbf{D}$ in the both the conditional and unconditional setting, along with the LOS Predictor $\mathcal{L}$.
\begin{table}
    \caption[NN Size] {Parameter Count of  $\mathbf{G}$, $\mathbf{D}$ and $\mathcal{L}$}
    \label{tab:nn_size}
    \centering
    \begin{tabular}{|P{3cm}|p{2.5cm}|p{2.5cm}|}
        \hline
         & WGAN-GP & CWGAN \\\hline
        Generator $\mathbf{G}$ & 1,069,568 &  1,328,468 \\\hline
        Critic $\mathbf{D}$ & 100,753 & 112,181 \\\hline
        LOS Predictor $\mathcal{L}$ & \multicolumn{2}{c|}{101,201} \\\hline
    \end{tabular}
\end{table}

In Algorithm~\ref{alg:WGAN_GP_training},~\ref{alg:FedGAN_training} and~\ref{alg:PCGAN_training}, we set $n_{\mathrm{d}}=5,~\beta=10,~\tau=0.01$ and $\gamma=0.00005$ \cite{gulrajani2017improved}\cite{arjovsky2017wasserstein}.
We utilize a minibatch size of $m = 200$ in all GAN training. For performing GCE, we utilize an Adam \cite{kingma2014adam} optimizer with a step size $\eta = 0.1$, $\lambda_{\mathrm{reg}} = 0.001$ and iteration count $100$. We also determined empirically that resetting the RMSProp optimizer for the critic at every training iteration improved the performance of Algorithm \ref{alg:WGAN_GP_training} (refer Appendix~\ref{subsec:reset_optimizer})\footnote{Code and data will be made publicly available at https://github.com/akashsdoshi96/ota-gan-mimo-ce}.

\subsection{Compressed Sensing (CS) Baselines} \label{subsec:baselines}
We compare the performance of GCE with a few standard CS baselines. It should be noted that the objective of this paper is to design an OTA GAN training procedure for CE, while using GCE to perform channel estimation. As described in Section~\ref{subsec:related}, there have been recent works in the deep generative domain \cite{arvinte2021deep,jalal2021robust} that have the potential to further improve the performance and robustness of CE using generative networks. Analyzing the performance and viability of such methods for mmWave MIMO CE is left for future work.

\vspace{2mm}
\noindent 
i) \textit{Orthogonal Matching Pursuit (OMP):} As described in \cite{mendez2016hybrid}, OMP minimizes $||\underline{\mathbf{H}_\mathrm{v}}||_0$ subject to $||\underline{\mathbf{y}} - \mathbf{A}_{\mathrm{sp}}\underline{\mathbf{H}_\mathrm{v}}||_2 \leq \sigma$. To prevent overfitting to the noise, OMP must be stopped\footnote{We set the maximum number of iterations to 200.} when the energy in the residual is less than $\sigma^2$.

\noindent
ii) \textit{EM-GM-AMP:} We utilize the Approximate Message Passing algorithm EM-GM-AMP \cite{vila2013expectation}, which accepts as input $\underline{\mathbf{y}}$ and $\mathbf{A}_{\mathrm{sp}}$, and recovers the channel estimate $\underline{\mathbf{H}_\mathrm{v}}$.

\vspace{2mm}
\noindent
Note that unlike the CS baselines, whose number of iterations depend on the SNR level, GCE is run for a constant number of iterations at all SNR without its performance being impacted.

\section{Results \& Discussion} \label{sec:results}
In this section, we will present and analyze the results for each GAN architecture presented in Section~\ref{sec:gan_arch} in sequence, before concluding with results from PCGAN. The performance metric used to assess the quality of the channel estimate will be the NMSE as defined in \eqref{eq:NMSE}. We will primarily be presenting two figures/tables for each architecture : GCE based NMSE as a function of i) GAN training iterations at a fixed SNR and ii) SNR for a given generative model. All figures that plot NMSE vs training iterations have been evaluated at an SNR of 15 dB and we will utilize $N_s = 16$, $N_p = 25$, $N_{bit,t} = 6$ and $N_{bit,r} = 2$, unless otherwise stated.

\subsection{WGAN-GP} \label{subsec:results_wgan_gp}
\begin{figure}
    \centering
    \subfloat[WGAN-GP for CDL A-E]{\includegraphics[width = 3.9in]{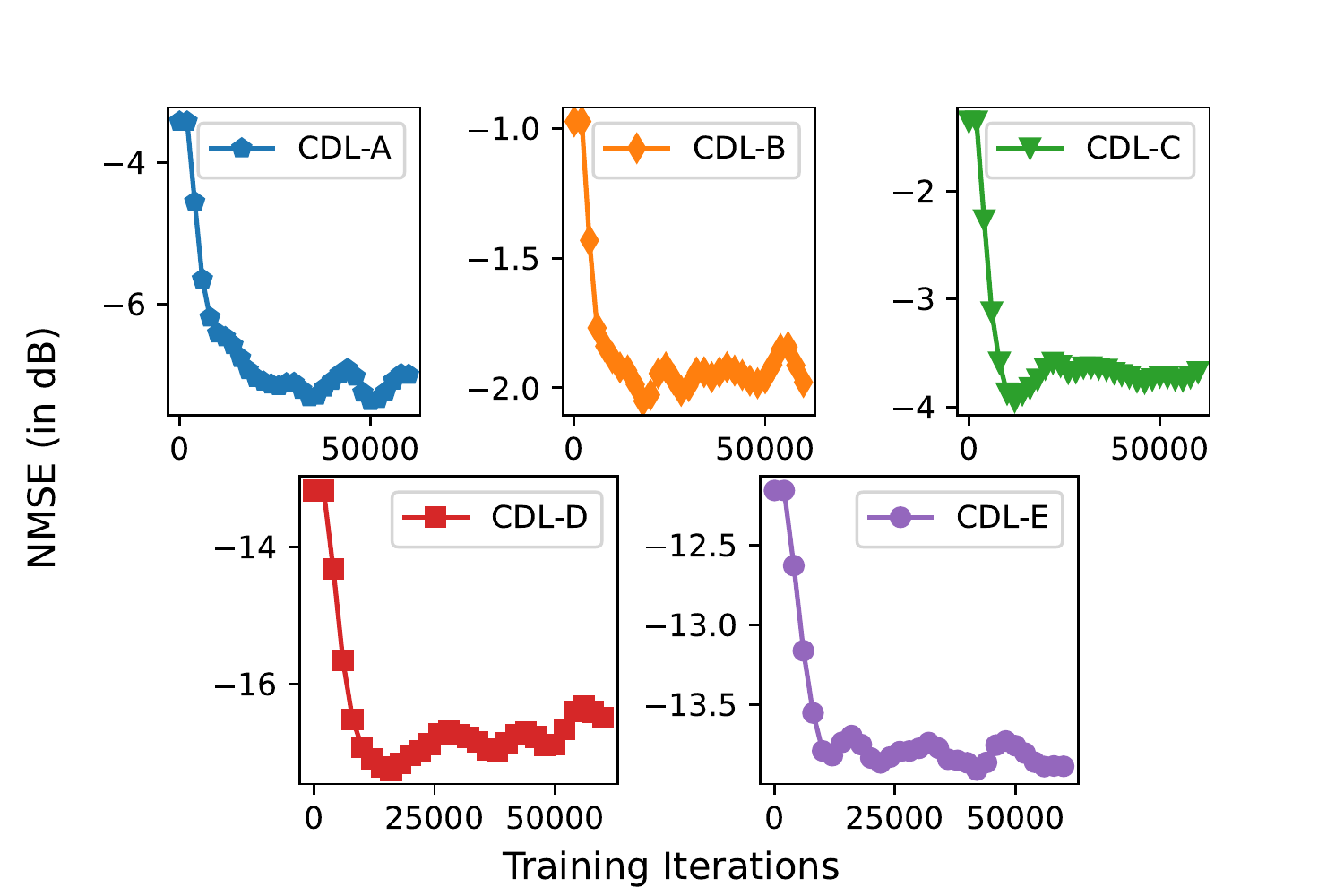} \label{fig:nmse_iter_wgan_gp}}
    \subfloat[WGAN vs WGAN-GP for CDL-A]{\raisebox{0.35in}{\includegraphics[width=2.5in]{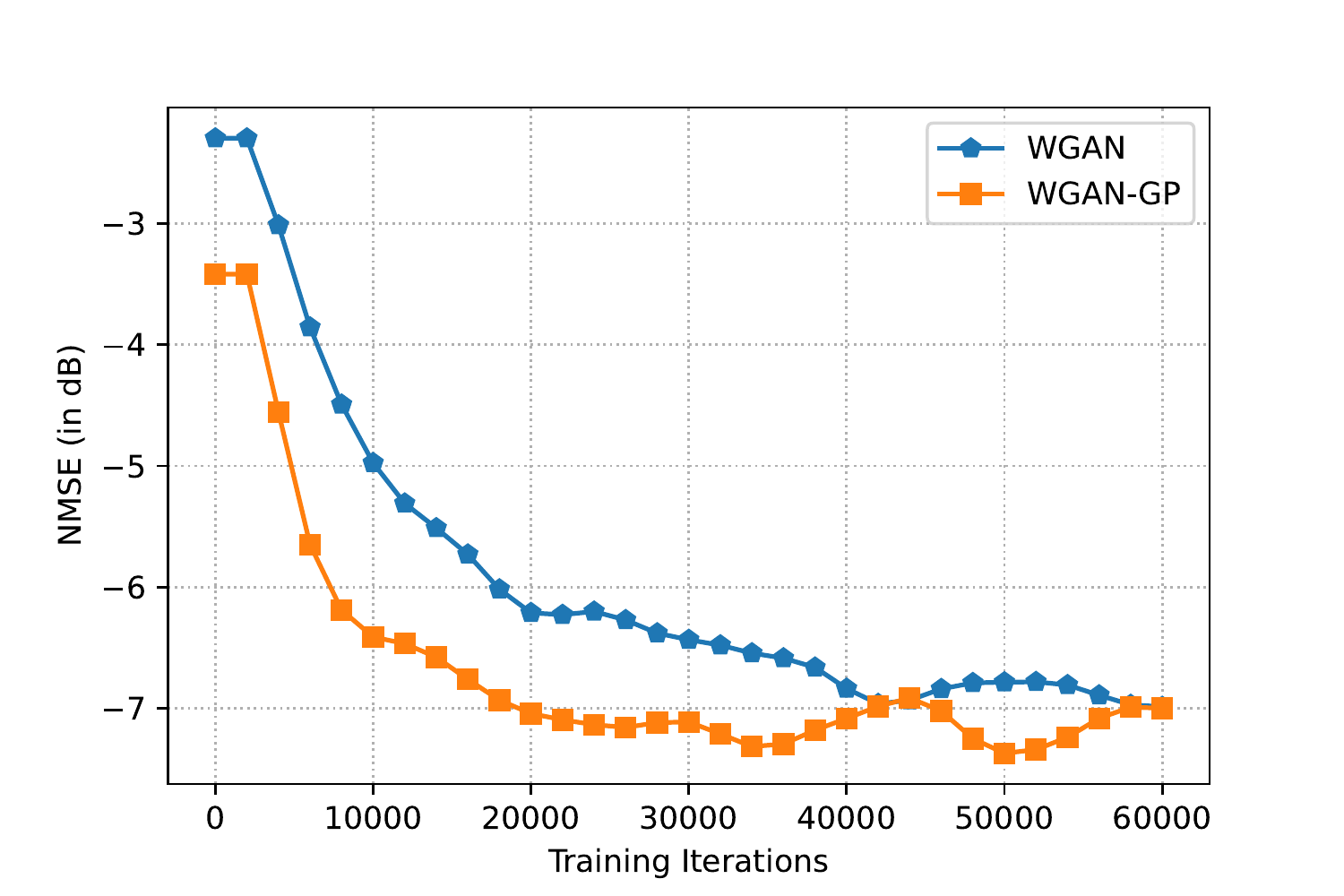}}\label{fig:nmse_iter_cdl_a_gp_nogp}}
    \caption{NMSE vs Training Iterations. NMSE increases with decrease in approximate beamspace sparsity.}
    \label{fig:nmse_iter_wgan}
\end{figure}

\begin{figure}
    \centering
    \includegraphics[width=4.5in]{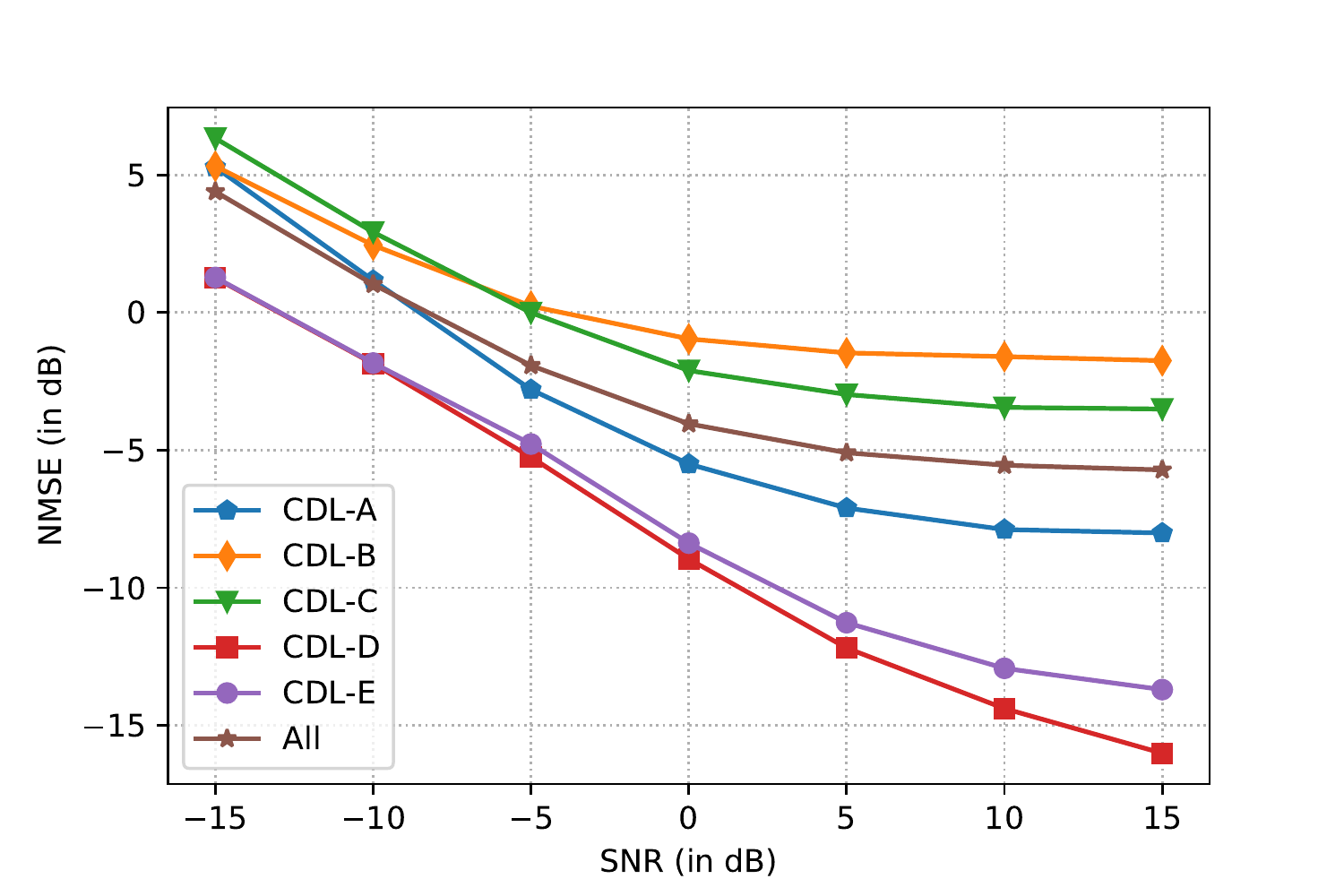}
    \caption{NMSE vs SNR for WGAN-GP for CDL A-E. ``All" denotes an average over all channel models.}
    \label{fig:nmse_snr_wgan_gp}
\end{figure}

We design a separate generator for each of the five CDL channel models by training a WGAN-GP ($\mathbbm{1}_{\mathrm{GP}} = 1$) using Algorithm~\ref{alg:WGAN_GP_training} for 60000 training iterations. The smoothed\footnote{Figures have occasionally been smoothed using a Hanning window of size 6 to improve readability, as and where mentioned.} plot of NMSE vs training iterations is shown in Fig.~\ref{fig:nmse_iter_wgan_gp}. We have also highlighted the gain of using WGAN-GP over WGAN in Fig.~\ref{fig:nmse_iter_cdl_a_gp_nogp}.
Using Fig.~\ref{fig:nmse_iter_wgan_gp}, we extract the trained generator with the lowest NMSE for each channel model (refer Remark~\ref{remark:nmse_train}) and plot NMSE vs SNR in Fig.~\ref{fig:nmse_snr_wgan_gp}. 
Furthermore, in Table~\ref{tab:gce_vs_cs}, we have compared the GCE based NMSE with the baselines outlined in Section~\ref{subsec:baselines} at three distinct SNR $=\{-5,0,10\}$ dB\footnote{SNR $=-5$ dB is a commonly utilized minimum SNR threshold for initial access \cite{giordani2016initial}.}.
\begin{table}
    \caption[GCE vs CS] {NMSE (in dB) of WGAN-GP based GCE and CS Baselines for CDL A-E}
    \label{tab:gce_vs_cs}
    \centering
    \begin{tabular}{|c|c|c|c|c|c|c|}
        \hline
        \multicolumn{2}{|c|}{}& CDL-A & CDL-B & CDL-C & CDL-D & CDL-E\\\hline
        \multirow{3}{6.3em}{SNR = $-5$ dB} & GCE & \textbf{-2.80} & \textbf{0.23} & \textbf{-0.01} & \textbf{-5.25} & \textbf{-4.78} \\
        & OMP & 2.26 & 2.31 & 2.28 & 0.48 & 1.82\\
        & EM-GM-AMP & 0.35 & 0.59 & 0.52 & -1.08 & -0.09\\\hline
        \multirow{3}{6.4em}{SNR = $0$ dB} & GCE & \textbf{-5.50} & \textbf{-0.95} & \textbf{-2.11} & \textbf{-8.95} & \textbf{-8.38} \\
        & OMP & 1.22 & 1.39 & 1.31 & -1.76 & -0.25\\
        & EM-GM-AMP & -0.67 & 0.14 & 0.04 & -2.70 & -1.71\\\hline
        \multirow{3}{6.3em}{SNR = $10$ dB} & GCE & \textbf{-7.88} & \textbf{-1.60} & \textbf{-3.45} & \textbf{-14.38} & \textbf{-12.92} \\
        & OMP & 0.73 & 0.64 & 0.55 & -3.26 & -2.21\\
        & EM-GM-AMP & -1.83 & -1.55 & -1.63 & -2.80 & -2.61\\\hline
    \end{tabular}
\end{table}

We observe that across CDL channel models, the performance of GCE is consistently $\mathrm{B} < \mathrm{C} < \mathrm{A} < \mathrm{E} < \mathrm{D}$. This is in agreement with the decreasing number of rays/clusters and the increasing magnitude of the LOS component in $\mathbf{H}_{\mathrm{v}}$ as we go from left to right (refer Table 7.7.1 of \cite{3gpp.38.901} for the precise channel profiles). Moreover, GCE consistently outperforms OMP, while also outperforming EM-GM-AMP for all CDL models excluding CDL-B, for which the performance of GCE and EM-GM-AMP are approximately similar. GCE also significantly outperforms both baselines at all SNR for LOS channels -- CDL-D and CDL-E.  

\begin{remark} \label{remark:nmse_train}
\textit{All WGAN models designed in Section~\ref{sec:gan_arch} are trained to directly minimize $L(\theta_d)$ and maximize $-L(\theta_g)$, \textbf{not} to minimize NMSE. While \cite{arjovsky2017wasserstein} has noted that the learning curves of a WGAN are empirically seen to correlate well with observed sample quality, this does not imply that the NMSE will decrease smoothly with training iterations as will be evident in most NMSE vs training iteration plots in this paper. NMSE convergence smoothness will be further degraded when we use noisy data in Pilot GAN and PCGAN in Section~\ref{subsec:results_pilot_gan} and~\ref{subsec:results_pcgan} respectively.}
\end{remark}

\vspace{2mm}
\noindent
\textbf{Impact of mutual coherence of the sensing matrix $\mu(\mathbf{A}_{\mathrm{sp}})$:} Consider the standard CS problem -- determine $\mathbf{x}$ given $\mathbf{y} = \mathbf{Ax} + \mathbf{n}$ with $\mathbf{A} \in \mathbb{R}^{m \times n}$ and $m < n$. Suppose $\mathbf{A} = [\mathbf{a}_1, \ldots \mathbf{a}_n]$, then the mutual coherence of $\mathbf{A}$ is given by
\begin{equation*}
    \mu(\mathbf{A}) = \underset{i,j; i \neq j}{\rm max} \frac{|\mathbf{a}_{i}^H\mathbf{a}_{j}|}{||\mathbf{a}_{i}||_2||\mathbf{a}_{j}||_2}.
\end{equation*}
Compressed sensing using generative models \cite{bora2017compressed} \cite{jalal2020robust} typically employs measurement matrices $\mathbf{A}$ whose entries are i.i.d. and chosen from simple random distributions such as Gaussian or sub-Gaussian. Consequently, it is highly likely that $\mathbf{A}$ is a full-rank matrix with low cross correlation among its columns.
In other words, $\mathrm{rank}(\mathbf{A}) = m$ with high probability, and the mutual coherence $\mu(\mathbf{A})$ will be low.
For a given $\mu(\mathbf{A})$, if $\mathbf{x}$ satisfies $||\mathbf{x}||_0 < \frac{1}{2}\big(\frac{1}{\mu(\mathbf{A})} + 1\big)$, then there exists a unique $\mathbf{x}$ with the given $\mathcal{L}_0$ norm that satisfies $\mathbf{y} = \mathbf{Ax}$ \cite{duarte2011structured}. Hence low mutual coherence decreases the reconstruction error of $\mathbf{x}$. Owing to the low $\mu(\mathbf{A})$, we believe that \cite{bora2017compressed} fails to truly convey the advantage of using generative models for compressed sensing over other traditional CS baselines for CE.

\begin{figure}
    \centering
    \subfloat[$\mu(\mathbf{A}_{\mathrm{sp}})$ and $\mathrm{rank}(\mathbf{A}_{\mathrm{sp}})$ vs $N_s$]{\includegraphics[width=3.2in]{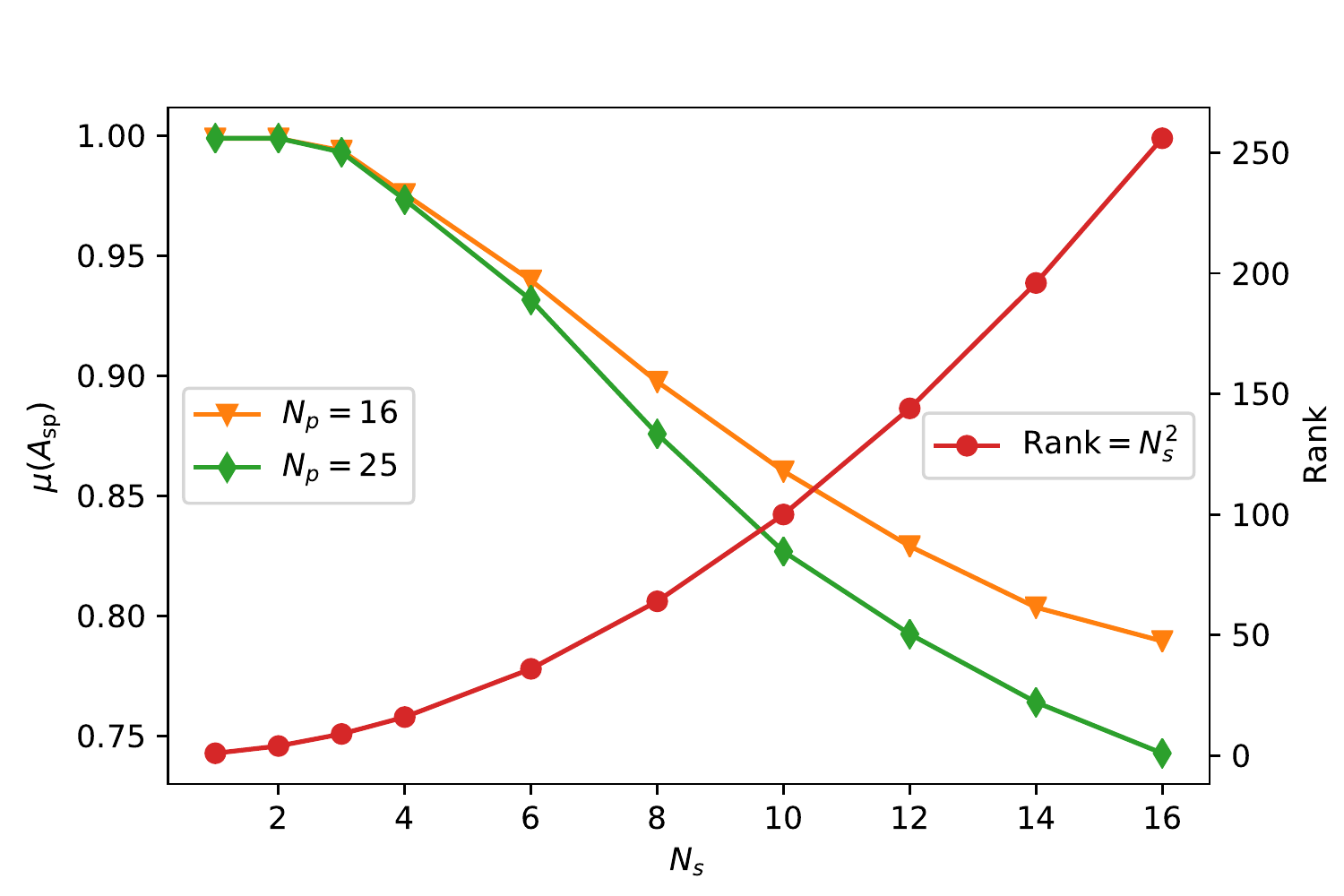}\label{fig:mu_rank_N_s}}
    \hspace{0.04in}
    \subfloat[NMSE vs $N_s$]{\includegraphics[width=3.2in]{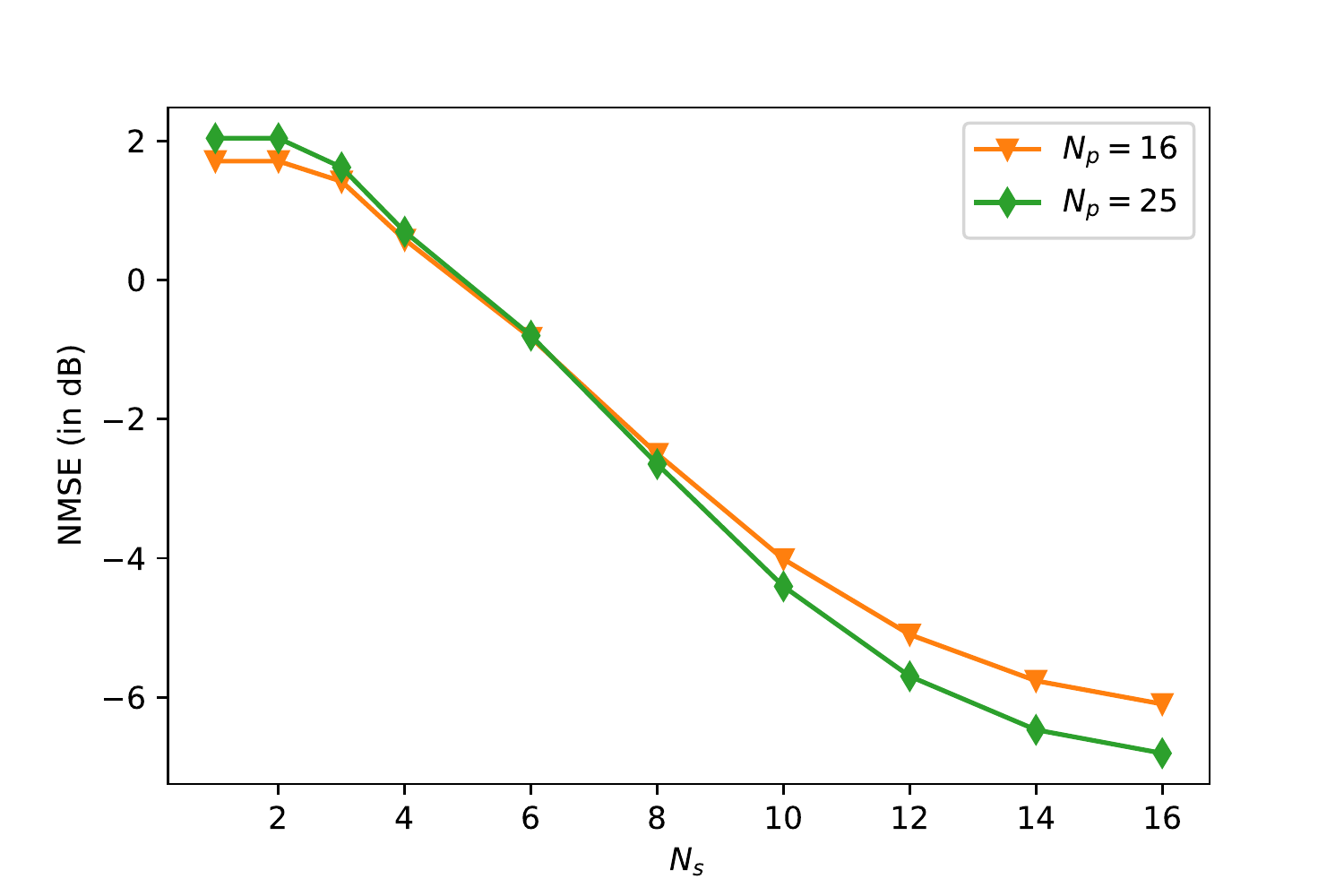}\label{fig:nmse_N_s}}
    \caption{NMSE, $\mu(\mathbf{A}_{\mathrm{sp}})$ and $\mathrm{rank}(\mathbf{A}_{\mathrm{sp}})$ vs $N_s$ for CDL-A at SNR $= 15$ dB. Low $\mu(\mathbf{A}_{\mathrm{sp}})$ decreases NMSE.}
    \label{fig:nmse_mu_rank_N_s}
\end{figure}

In Section~\ref{sec:gce}, we defined the measurement matrix as $\mathbf{A}_{\mathrm{sp}} = (\mathbf{A}_{\mathrm{T}}^H\mathbf{F}\mathbf{S})^T \otimes \mathbf{W}^H\mathbf{A}_{\mathrm{R}}$, with
$\mathrm{rank}(\mathbf{A}_{\mathrm{sp}}) \leq N_s^2$. By varying $N_s$ from $1$ to $16$ for $N_p = 16$ and $N_p = 25$, we observe that $\mu(\mathbf{A}_{\mathrm{sp}}) \geq 0.75$ in Fig.~\ref{fig:mu_rank_N_s}. This implies that a unique $\mathbf{H}_{\mathrm{v}}$ can be recovered by OMP only if $||\mathbf{H}_{\mathrm{v}}||_0 < 2$.  This is consistent with our observations in Table~\ref{tab:gce_vs_cs}, where OMP performs quite poorly. Meanwhile, GCE significantly outperforms OMP, showing that the rich generative prior compensates for the high mutual coherence of $\mathbf{A}_{\mathrm{sp}}$. Thus, GCE eliminates the need for careful tuning of the entries of $\mathbf{A}_{\mathrm{sp}}$ to minimize $\mu(\mathbf{A}_{\mathrm{sp}})$ as is done in \cite{mendez2016hybrid}.

However, the performance of GCE also improves if we improve the mutual coherence. In Fig.~\ref{fig:nmse_N_s}, we can observe that for the same rank at $N_s = 16$, when the value of $\mu(\mathbf{A}_{\mathrm{sp}})$ is higher for $N_p = 16$ than $N_p = 25$, GCE does attain a lower NMSE with $N_p = 25$. 

\subsection{CWGAN} \label{subsec:results_cwgan}
\begin{figure}
    \centering
    \subfloat[NMSE vs Training Iterations]{\includegraphics[width=3.2in]{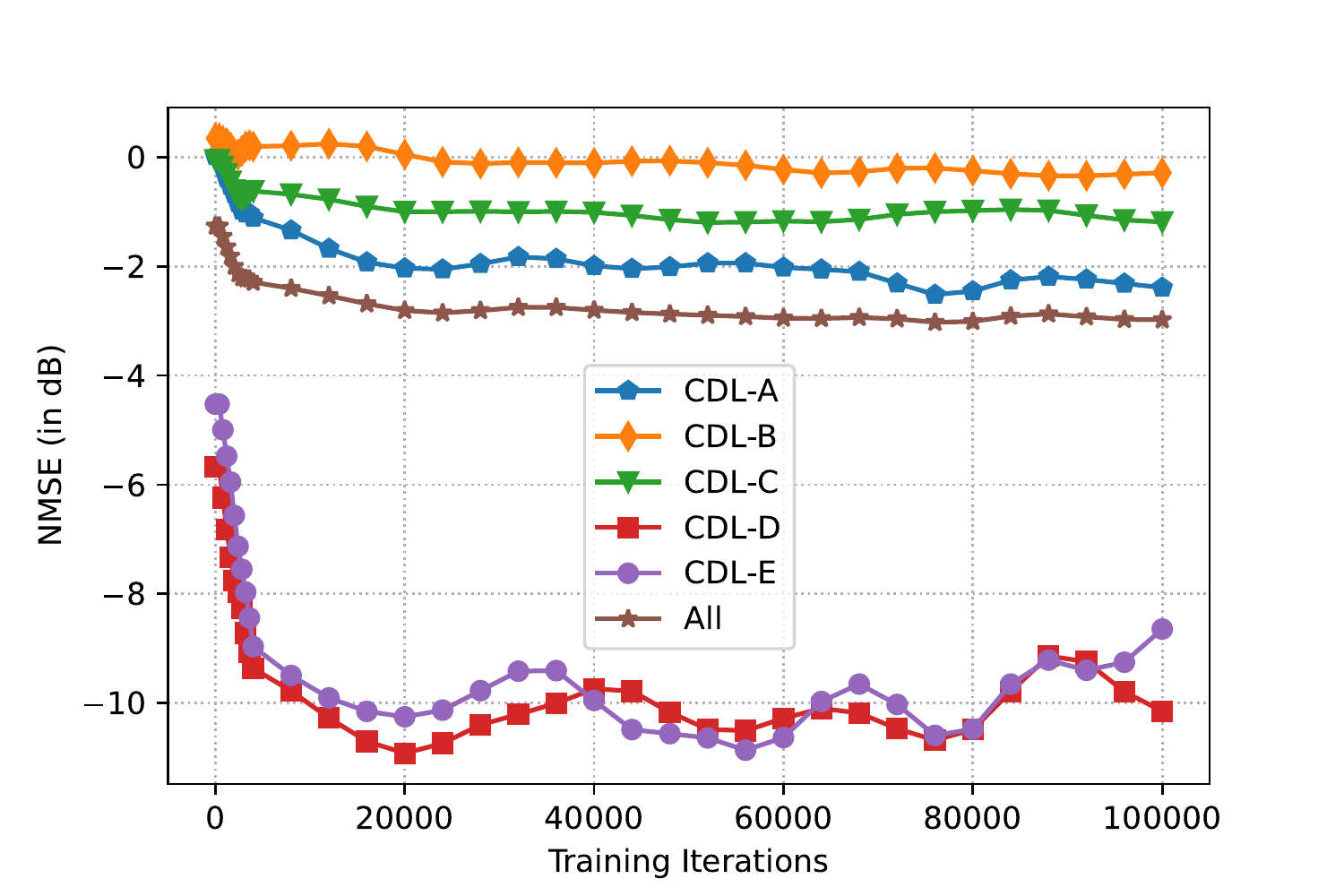}\label{fig:nmse_iter_cwgan_gp}}
    \hspace{0.04in}
    \subfloat[NMSE vs SNR]{\includegraphics[width=3.2in]{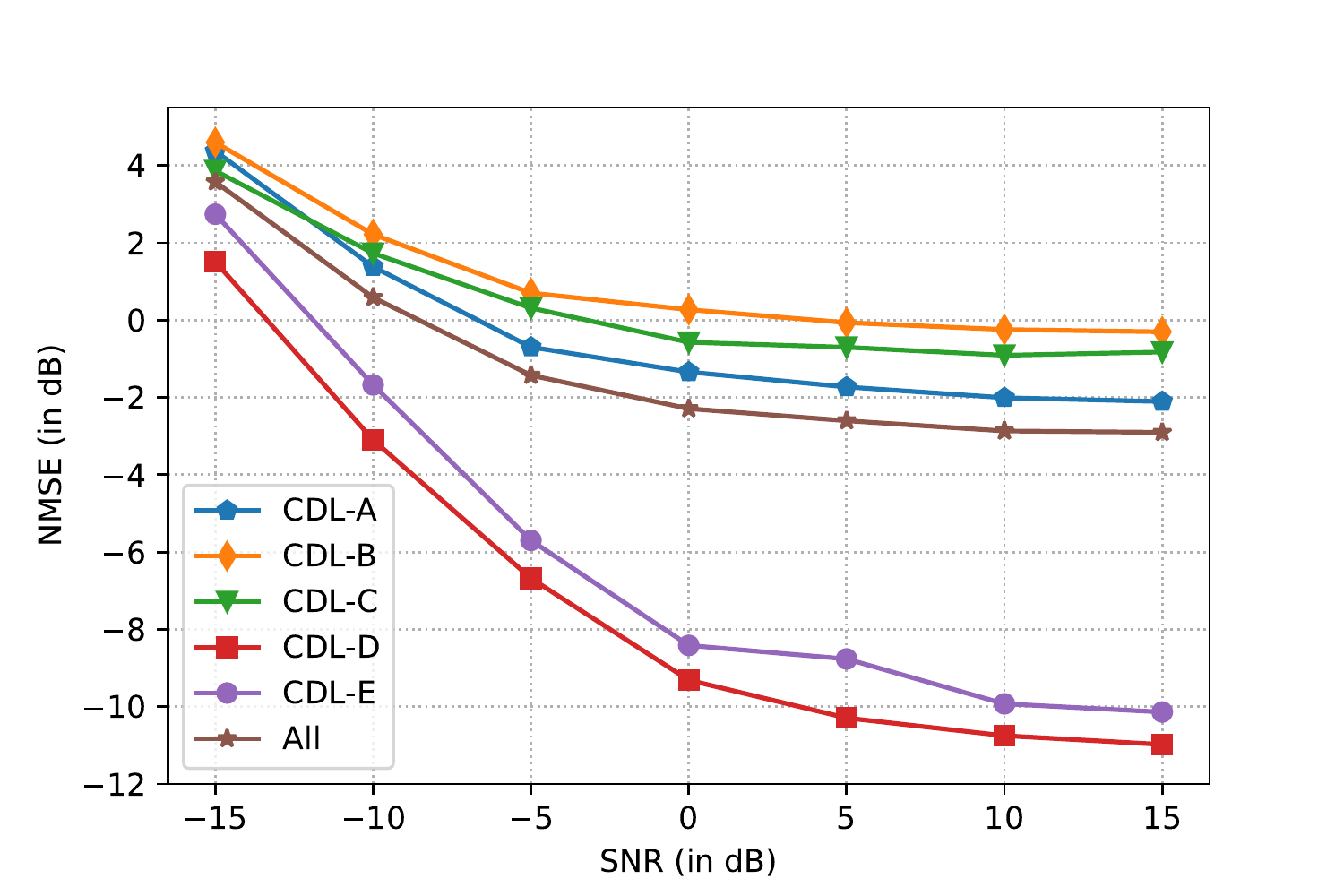}\label{fig:nmse_snr_cwgan_gp}}
    \caption{NMSE for CWGAN trained on CDL A-E. ``All" denotes an average over all channel models.}
    \label{fig:nmse_cwgan_gp}
\end{figure}

We now design a single conditional generative model by training a CWGAN ($\mathbbm{1}_{\mathrm{GP}} = 0$) as outlined in Section~\ref{subsec:cwgan_gp} for 100,000 training iterations. For CWGAN, we assume that the LOS/NLOS ground truth label $\chi$ is known while performing GCE. The smoothed plot of NMSE vs training iterations is shown in Fig.~\ref{fig:nmse_iter_cwgan_gp}, and the plot of NMSE vs SNR for the best generative model (determined using Fig.~\ref{fig:nmse_iter_cwgan_gp}) is shown in Fig.~\ref{fig:nmse_snr_cwgan_gp}. The degradation in average NMSE compared to the individually trained WGAN-GP models increases from $0.5$ dB at $-5$ dB SNR to $1.7$ dB at $0$ dB SNR to $2.6$ dB at $10$ dB SNR. However, GCE using CWGAN continues to outperform OMP at all the aforementioned SNR for CDL A-E, and even outperforms EM-GM-AMP for CDL-A, D and E, with the gain being largest for LOS channel models ($\sim 6$ dB) and $\sim 0.7$ dB for CDL-A.

The reduction in gain on switching to a conditional model can be attributed to the simple binary condition being utilized. For instance, CDL-A and CDL-B are both NLOS channel models, but have significantly different beamspace representations. We are actively investigating the possibility of introducing a learnable condition into the GAN framework such that the condition more accurately reflects the estimated degree of approximate sparsity in $\mathbf{H}_\mathrm{v}$.

\subsection{Pilot GAN} \label{subsec:results_pilot_gan}
We consider two cases, $U = 1$ and $U = 4$, where $U$ is the number of UEs associated with a BS, and utilize full-rank pilot measurements i.e. of rank $N_tN_r$ for training in both cases with $\mathbbm{1}_{\mathrm{GP}} = 1$.

\subsubsection{$U=1$} 
\begin{figure}
    \centering
    \subfloat[NMSE vs Training Iterations]{\includegraphics[width=3.2in]{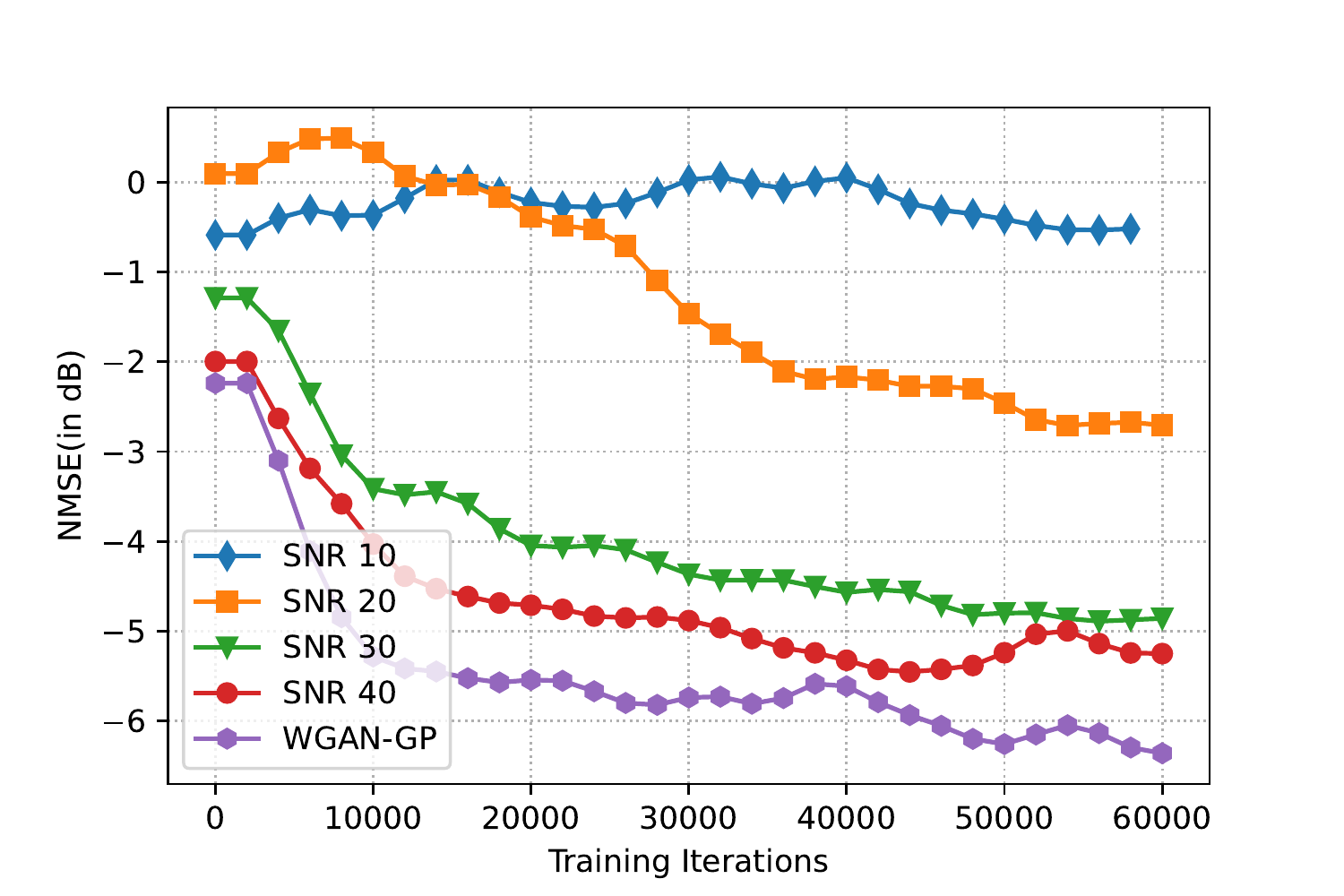}\label{fig:nmse_iter_pilot_gan}}
    \hspace{0.04in}
    \subfloat[NMSE vs SNR]{\includegraphics[width=3.2in]{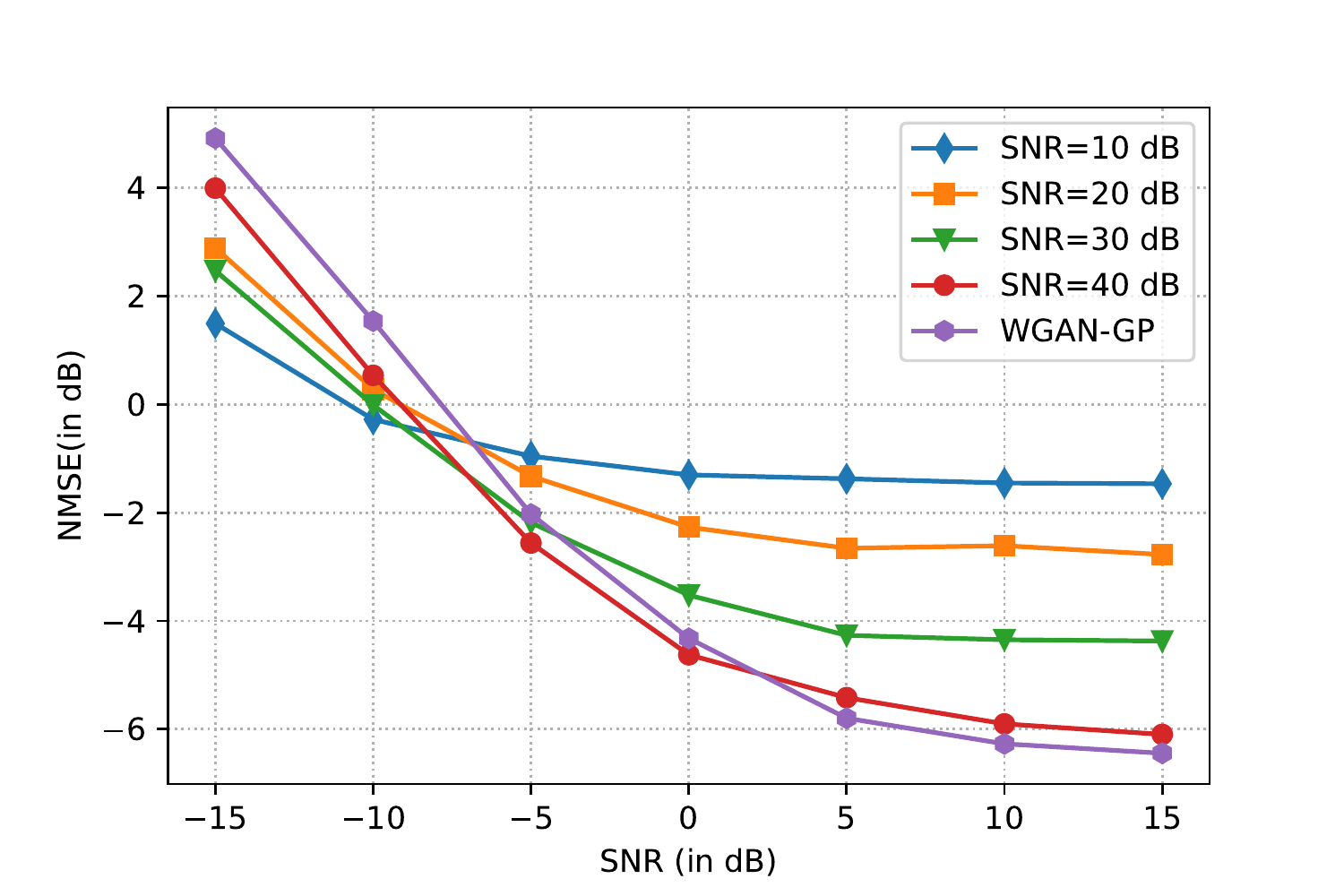}\label{fig:nmse_snr_pilot_gan}}
    \caption{NMSE for Pilot GAN trained on CDL-A. Note that the labels are \textit{training} SNR, while the x-axis in the right figure is the \textit{test} SNR. WGAN-GP denotes no noise during training. NMSE increases at lower \textit{training} SNR.}
    \label{fig:nmse_pilot_gan}
\end{figure}

We train Pilot GAN as outlined in Section~\ref{subsec:pilot_gan} for 60,000 training iterations. We investigate the noise tolerance of Pilot GAN by training a generative model for CDL-A at SNR = $\{10,20,30,40\}$ dB. The SNR value will determine the variance $\sigma^2$ of each term in $\mathbf{\underline{n}}[1:K]$ in \eqref{eq:pilot_gan_stack_sys_model} during training. The smoothed plot of NMSE vs training iterations is shown in Fig~\ref{fig:nmse_iter_pilot_gan}, and the plot of NMSE vs SNR in Fig.~\ref{fig:nmse_snr_pilot_gan}. From Fig.~\ref{fig:nmse_iter_pilot_gan}, we can see that the NMSE at a training SNR of $40$ dB is roughly similar to the noiseless case, while the NMSE convergence at a training SNR of $\leq 20$ dB is significantly degraded. However, the NMSE improves at SNR $\leq -10$ dB over WGAN-GP for all training SNR, where WGAN-GP in Fig.~\ref{fig:nmse_snr_pilot_gan} denotes training on clean channel realizations. This could be a possible advantage of training with noisy data.

While this might seem a high SNR threshold to meet, two points can be noted: i) Training a single generative model for a larger class of CDL channels could potentially reduce the SNR induced degradation, as we will see in Section~\ref{subsec:pcgan} and ii) Practical wireless deployments \cite{3gpp.38.214} use precoders and combiners attuned to channel realizations from a few time-slots ago, which yields significantly better link SNR than random precoders and combiners.

\subsubsection{$U = 4$}
\begin{figure}
    \centering
    \includegraphics[width=3.7in]{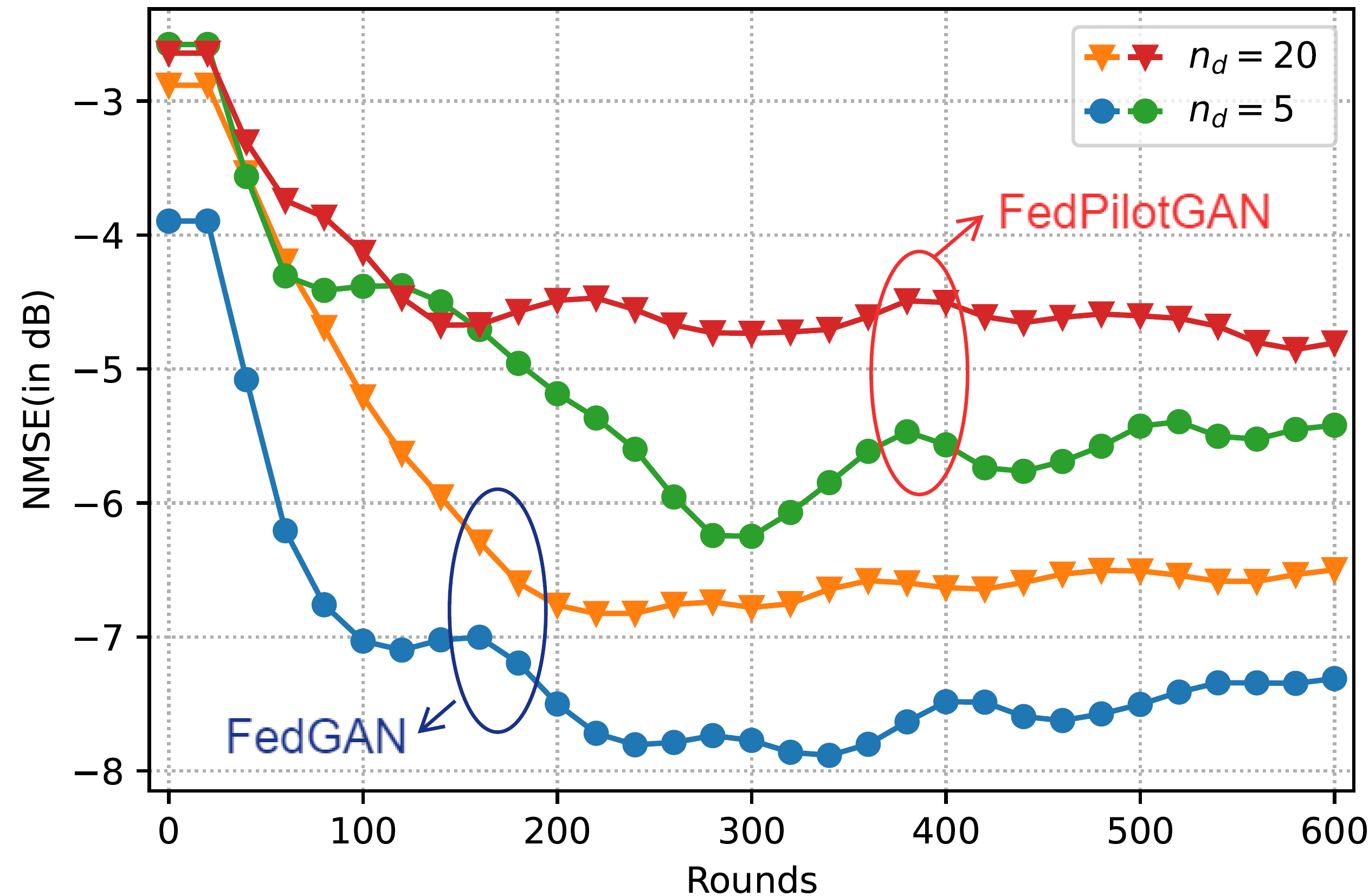}
    \caption{NMSE vs Rounds for Federated Pilot GAN training on CDL-A. Note the trade-off between BS-UE communication frequency and NMSE, as well as between training noise levels and NMSE.}
    \label{fig:fedgan}
\end{figure}
We train Federated Pilot GAN as outlined in Algorithm~\ref{alg:FedGAN_training} for 600 rounds on CDL-A, with $D = 500$ and $m=200$. Instead of fixing the link SNR, we perform a network level simulation using the parameters outlined in Table~\ref{tab:fedGAN}. We fix the large scale path loss for the duration of the simulation, and assume it is known at the UE. We investigate the impact of two factors on the NMSE convergence vs rounds -- i) noise in LS channel estimate and ii) BS-UE communication frequency. To analyze i), we consider perfect channel estimates i.e. $\mathbf{H}_{\mathrm{v},LS} = \mathbf{H}_{\mathrm{v}}$ for training the GAN, and refer to this as ``FedGAN", while the GAN trained on LS channel estimates generated using the parameters in Table \ref{tab:fedGAN} is referred to as ``FedPilotGAN". In order to study ii), we change the number of critic updates ($n_d$ in Algorithm \ref{alg:FedGAN_training}) while keeping the total number of training iterations equal to the centralized Pilot GAN implementation. We consider $n_d=5$ and $n_d=20$ which in turn implies $l=100$ and $l=25$ respectively. Note that $n_d=20$ also involves performing $4$ generative updates per training iteration instead of $1$, to have a fair comparison.

The smoothed plot of NMSE vs rounds is shown in Fig.~\ref{fig:fedgan}. From the performance of ``FedGAN" with $n_d=5$, as contrasted with CDL-A in Fig. \ref{fig:nmse_iter_wgan_gp}, we can ascertain that simply moving to a federated implementation did not have any detrimental impact on the NMSE achieved. However, noise in the LS channel estimates worsens the NMSE by $\sim 2$ dB. Furthermore, lowering the BS-UE communication frequency by a factor of $4$ worsens the NMSE by $\sim 0.8$ dB. Such a trade-off between performance and BS-UE communication frequency as well as between performance and noise in the training data are to be expected in a federated setting \cite{bonawitz2019towards}.
\begin{table}
    \centering
    \begin{tabular}{|c|c|}
        \hline
        BS Transmit Power & $23$ dBm  \\\hline
        Noise PSD & $-174$ dBm/Hz \\\hline 
        Bandwidth & $20$ MHz \\\hline
    \end{tabular}
    \hspace{0.02in}
    \begin{tabular}{|c|c|}
        \hline
        UE Noise Figure & 9 dB\\\hline
        Path Loss model & InH-Office \cite{3gpp.38.901} \\\hline
        Fading model & Block fading \\\hline
    \end{tabular}
    \caption{Federated Pilot GAN simulation parameters}
    \label{tab:fedGAN}
\end{table}

\vspace{2mm}
\noindent
\textbf{Applications of Federated Pilot GAN:} The applicability of FedPilotGAN is not restricted to building a generative channel estimator in an OTA fashion as is the focus in this paper. Here we describe a federated design of a (clean) data generator from noisy data realizations. Consider a BS (server) that requires a large dataset of channels (images) for some task. The BS (server) does not possess the data itself, however it is connected to a set of UEs (clients) that can capture noisy\footnote{Obtaining multiple high resolution images can often be expensive and impractical for many sensing and tomography problems such as MRI and astronomical imaging \cite{bora2018ambientgan} \cite{jalal2020robust}.} copies of channel (image) realizations \cite{li2014scaling}\cite{hardy2019md}. From a practical perspective, a UE cannot store large amounts of data, cannot perform computationally intensive training or tolerate the uplink overhead of communicating ``denoised" channels (images).  Using the FedPilotGAN framework, we can train a generative model at the BS, without being subject to any of the aforementioned disadvantages, that can produce channel (image) samples from the underlying distribution. In addition, having a generator at the BS eliminates the need for storing large amounts of data at the BS as well i.e. we can simply generate batches of the required size on the fly by sampling from the generator.

\subsection{PCGAN with LOS Predictor} \label{subsec:results_pcgan}
\subsubsection{LOS Predictor} 
\begin{figure}
    \centering
    \includegraphics[width=3.2in]{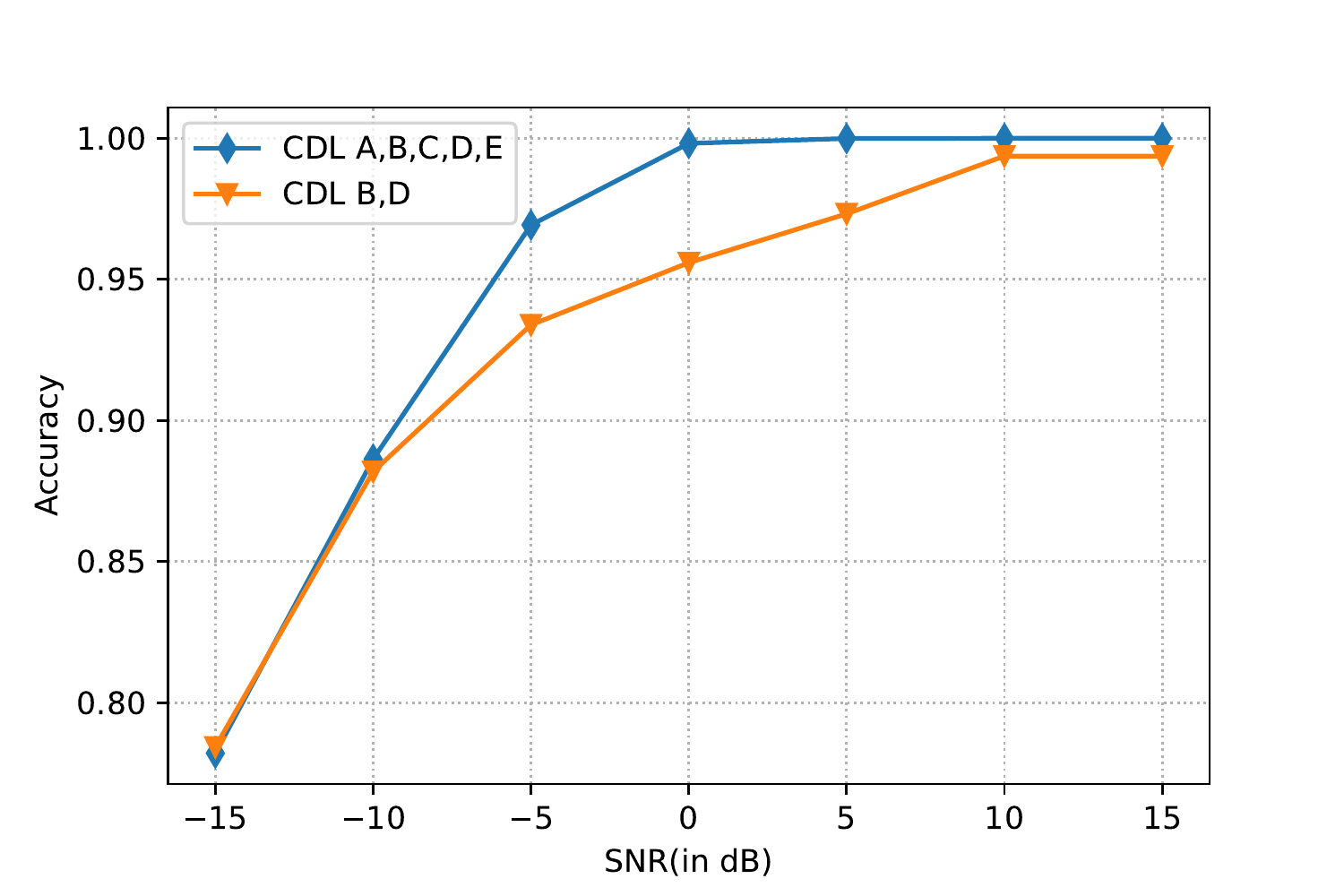}
    \caption{Accuracy of LOS Predictor on CDL A-E. Using a subset of channel models for training minimally impacts accuracy.}
    \label{fig:los_predictor}
\end{figure}

We train the LOS Predictor $\mathcal{L}(.;\theta_L)$ for 20,000 iterations using a batch size of $200$ and Adam \cite{kingma2014adam} optimizer with $\gamma = 0.0003$. The accuracy of $\chi$ as given by \eqref{eq:cond_LOS} as a function of SNR has been plotted in Fig.~\ref{fig:los_predictor}. Two curves have been plotted - the upper curve is when $\mathcal{L}(.;\theta_L)$ is trained and tested on all CDL channel models, and the lower curve is when $\mathcal{L}(.;\theta_L)$ is trained on CDL-B and CDL-D only and tested on CDL A-E. The relatively minimal loss in accuracy supports supervised training of an LOS Predictor. We can simply train it using channel realizations from simulation tools available offline and test it with pilot measurements that need not be from the same channel distribution.

\subsubsection{PCGAN}
\begin{figure}
    \centering
    \includegraphics[width=6.5in]{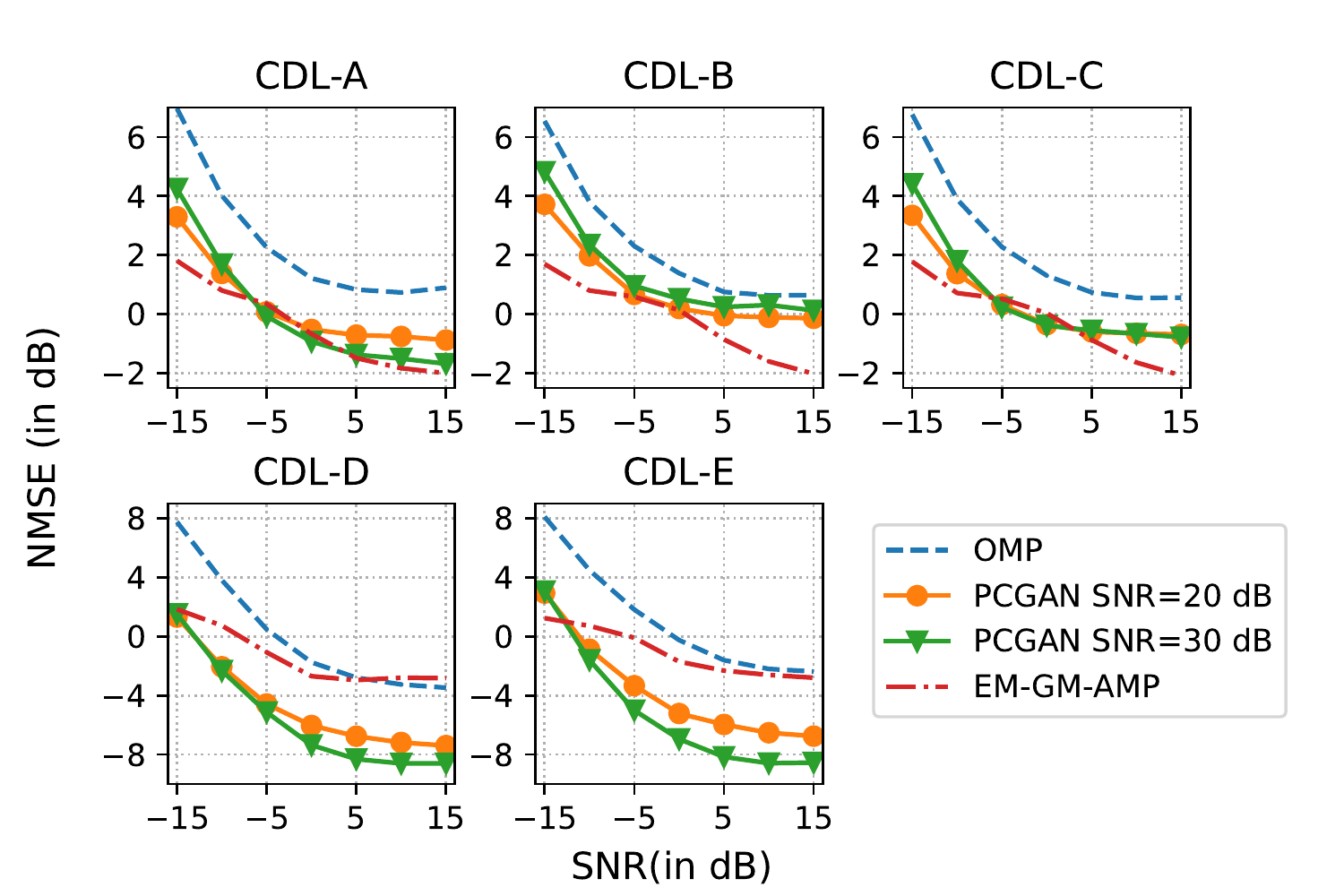}
    \caption{NMSE vs SNR for PCGAN at \textit{training} SNR of 20 and 30 dB. For reference, OMP and EM-GM-AMP baselines have also been plotted. PCGAN outperforms on LOS models, and achieves competitive results on NLOS models as compared to EM-GM-AMP, while consistently outperforming OMP.}
    \label{fig:nmse_snr_pcgan}
\end{figure}

We now use the trained $\mathcal{L}(.;\theta_L)$ to compute $\chi$ from the LS channel estimate $\mathbf{H}_{\mathrm{v},LS}$ of the received pilot measurements $\mathbf{\underline{y}}[1:K]$, and subsequently utilize $\{\mathbf{H}_{\mathrm{v},LS}\}$ to design a conditional generator by training a PCGAN ($\mathbbm{1}_{\mathrm{GP}} = 1$) using Algorithm~\ref{alg:PCGAN_training} for 100,000 training iterations. The NMSE vs SNR on the best generative model for training SNR of 20 and 30 dB has been depicted in Fig. \ref{fig:nmse_snr_pcgan} for each of CDL A-E. It should be noted that we utilize $\mathcal{L}(.;\theta_L)$ only for \textit{training} PCGAN. NMSE computation during \textit{inference} bypasses the LOS Predictor by solving \eqref{eq:gce_cond} in place of \eqref{eq:gce}. 

The NMSE on the LOS channel models degrades by $\sim 2$ dB for training SNR $30$ dB and $\sim 3.5$ dB for training SNR $20$ dB compared to CWGAN. Meanwhile, the NMSE on the NLOS channel models CDL-B and CDL-C is practically unaffected compared to CWGAN, and more interestingly, a better model has been learnt at the lower $20$ dB training SNR for CDL-B. PCGAN based GCE continues to outperform OMP for both training SNR, while outperforming EM-GM-AMP for the LOS channel models and being similar for CDL-A. 

\subsection{Complexity Analysis}
In our prior work \cite{balevi2020high}, we analyzed the computational complexity of GCE and contrasted it with OMP and EM-GM-AMP. Note that the complexity of both a forward and backward pass through a CNN are $\mathcal{O}(N_tN_r)$ -- backward propagation takes roughly twice the time of forward propagation for a CNN \cite{He2015Sun} and the complexity of a forward pass through $\mathbf{G}$ or $\mathbf{D}$ is $\mathcal{O}(N_tN_r)$ \cite{balevi2020high}. Given that each NN based operation in Algorithm~\ref{alg:PCGAN_training} can be expressed as either a forward or backward pass through a CNN, and the remaining are matrix additions, the big-$\mathcal{O}$ complexity of one iteration of PCGAN, given the dataset $\{\mathbf{H}_{\mathrm{v},LS}\}$, is $\mathcal{O}(N_tN_r)$. However, such an analysis has its limitations -- i) it does not account for the batch size $m$, which is practically constrained by the parallel compute capacity of a GPU and ii) the big-$\mathcal{O}$ complexity simply upper bounds a forward pass by $\mathcal{O}(N_tN_r)$, while the actual complexity may be significantly lower.

\section{Conclusions \& Future Directions} \label{sec:conclusion}
In this paper, we have presented a novel OTA formulation for training a GAN to learn the MIMO channel distribution for a range of LOS and NLOS channel models in accordance with the 3GPP standard \cite{3gpp.38.901}. We designed a LOS Predictor to determine whether the received pilot measurements originated from a LOS or a NLOS channel, and used the conditional output as input to a generative model, while the critic in the GAN was trained to distinguish between real and generated LS channel estimates. Subsequently, we utilized the trained generative model to perform generative channel estimation in the beamspace domain from compressive pilot measurements. While the individually trained WGANs outperformed both CS baselines -- OMP and EM-GM-AMP -- PCGAN was able to outperform OMP while outperforming EM-GM-AMP on LOS channel models and achieving competitive results on NLOS channels. 

As part of future work, channel estimation using generative networks could potentially be performed in the presence of transmit non-linearities as well using the robust median-of-means estimator designed in \cite{jalal2020robust}. There are also several research directions one could pursue to improve the practicality and applicability of the GAN training framework designed in this paper. In order to remove the assumption of block fading, we would need to train recurrent GANs \cite{esteban2017real} that produce a sequence of temporally correlated channel realizations. As explained in Section~\ref{subsec:results_cwgan}, the LOS predictor does not provide a sufficiently accurate indicator of the estimated degree of sparsity in the channel being reconstructed. Hence, we could instead train 3 GANs -- for low, medium and high levels of beamspace sparsity -- and then learn a classifier that will indicate which generative model to utilize \cite{ParkYBCP18}. Furthermore, we have not considered the degradation owing to quantization of the received signal, and incorporating CS based techniques for channel estimation from quantized pilot measurements \cite{myers2020low} in combination with generative models \cite{Qiu19WeiQiu} is an exciting future direction. Finally, one could look into combining GCE and GAN training to remove the requirement of full-rank pilot measurements for training Pilot GAN \cite{kabkab2018task}.

\section{Acknowledgements}
The authors would like to thank Chris Dick at NVIDIA for arranging the donation of a DGX Station A100 workstation to accelerate the training carried out in this work, and his feedback on this work.

\bibliographystyle{IEEEtran}
\bibliography{bibtex.bib}
\newpage

\appendix
\subsection{Proof of Section~\ref{subsec:pilot_gan}} \label{subsec:proof}
Consider noiseless pilot measurements $\mathbf{\underline{y}} = \mathbf{A}_{\mathrm{sp}}\underline{\mathbf{H}_\mathrm{v}}$ such that $N_pN_s < N_tN_r$. As explained in Section~\ref{sec:sys_model}, this implies that we cannot recover a unique $\underline{\mathbf{H}_\mathrm{v}}$ given only $\mathbf{\underline{y}}$. Let us utilize the LQ decomposition of $\mathbf{A}_{\mathrm{sp}}$ to represent it as $\mathbf{A}_{\mathrm{sp}} = \big(\mathbf{L} ~ \mathbf{0}\big) \mathbf{Q}$ where $\mathbf{L}$ is a $N_pN_s \times N_pN_s$ lower triangular matrix and $\mathbf{Q}$ is an $N_tN_r \times N_tN_r$ unitary matrix. Hence we have 
\begin{equation}
    \mathbf{\underline{y}} = \big(\mathbf{L} ~ \mathbf{0}\big) \mathbf{Q} \underline{\mathbf{H}_\mathrm{v}} = \big(\mathbf{L} ~ \mathbf{0}\big) \begin{bmatrix}
    \mathbf{h}_{Q,1}\\
    \mathbf{h}_{Q,2}
    \end{bmatrix}.
\end{equation}
where $\mathbf{h}_{Q,1}$ contains the first $N_pN_s $ entries of $\mathbf{Q} \underline{\mathbf{H}_\mathrm{v}}$ and $\mathbf{h}_{Q,2}$ contains the remaining. Denote $\underline{\mathbf{H}_{\mathrm{v},Q}} = \mathbf{Q} \underline{\mathbf{H}_\mathrm{v}}$. Since $\mathbf{Q}$ is invertible, there is a one-to-one mapping between the $p_{\underline{\mathbf{H}_{\mathrm{v},Q}}}(.)$ and $p_{\underline{\mathbf{H}_\mathrm{v}}}(.)$. However, $\underline{\mathbf{y}}$ is only a function of $\mathbf{h}_{Q,1}$. In fact, since $\mathbf{L}$ is lower triangular, we can compute $\mathbf{h}_{Q,1}$ from $\underline{\mathbf{y}}$ by elementary row operations. Hence all pdfs of the form
\begin{equation} \label{eq:pdf_transform}
    p_{\underline{\mathbf{H}_{\mathrm{v},Q}}}(.) = \frac{p_{\mathbf{\underline{y}}}(.)}{|\mathrm{det}(J_{\mathbf{\underline{y}}}(\mathbf{h}_{Q,1}))|} p_{\mathbf{h}_{Q,2}}(.),
\end{equation}
where $J_{\mathbf{\underline{y}}}(\mathbf{h}_{Q,1})$ is the Jacobian of $\mathbf{\underline{y}}$ evaluated at $\mathbf{h}_{Q,1}$, map to the same  $p_{\mathbf{\underline{y}}}(.)$. Note that \eqref{eq:pdf_transform} may not even include the ground truth $p_{\underline{\mathbf{H}_{\mathrm{v},Q}}}$ if $p_{\underline{\mathbf{H}_{\mathrm{v},Q}}} \neq p_{\mathbf{h}_{Q,1}}(.) p_{\mathbf{h}_{Q,2}}(.)$. Hence, we cannot define a one-to-one mapping between $p_{\mathbf{\underline{y}}}$ and $p_{\underline{\mathbf{H}_{\mathrm{v},Q}}}$, which implies the absence of the same between $p_{\mathbf{\underline{y}}}$ and $p_{\underline{\mathbf{H}_\mathrm{v}}}$, proving Ambient GAN \cite{bora2018ambientgan} and by extension -- Pilot GAN -- cannot be trained with compressive measurements.

\subsection{Optimal value of $d$ for $\mathbf{z} \in \mathbb{R}^d$} \label{subsec:optimal_d}
\begin{figure}
    \centering
    \subfloat[CDL-B as a function of $d$]{ \includegraphics[width=3.2in]{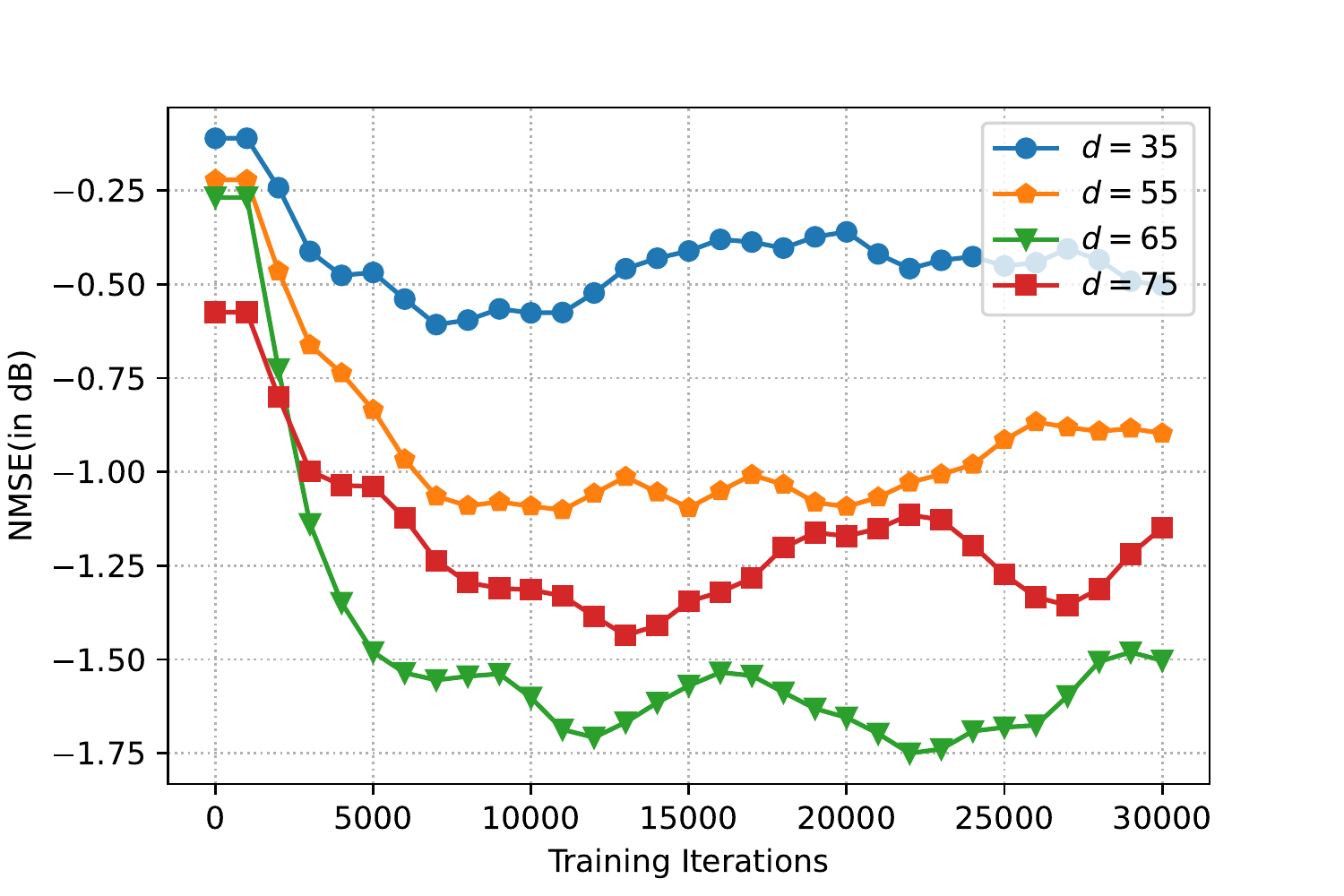}\label{fig:nmse_iter_d}}
    \subfloat[Impact of resetting critic optimizer (RO) on CDL-A]{\includegraphics[width=3.2in]{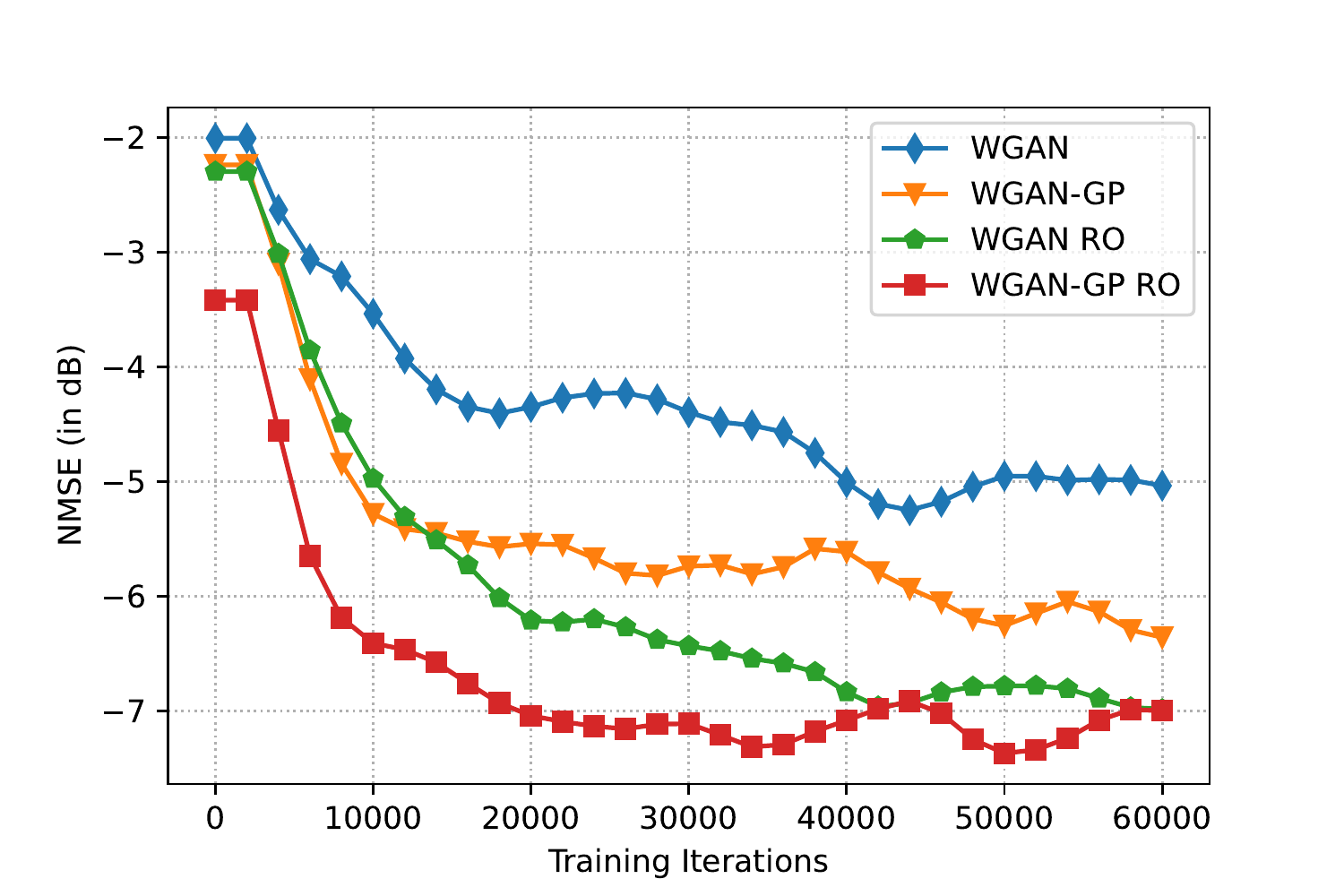}\label{fig:nmse_iter_ro}}
    \caption{NMSE vs Training Iterations for WGAN}
    \label{fig:nmse_iter_appendix}
\end{figure}

Given that 3GPP TR 38.901 \cite{3gpp.38.901} defines CDL-B to have the largest number of rays/clusters among all CDL models with only a weak LOS component, and consequently has the highest NMSE, we determine the optimal value of $d$ by training a WGAN-GP on CDL-B. The smoothed plot of NMSE vs training iterations is shown in Fig.~\ref{fig:nmse_iter_d} for varying values of $d$. The NMSE decreases going from $d=55$ to $d=65$, but increases on going to $d=75$. Hence we have chosen $d=65$ as the optimal latent dimension.

\subsection{Effect of resetting critic optimizer} \label{subsec:reset_optimizer}
Training instability in GANs can also be attributed to early critic convergence, which reduces the incentive for the generator to improve its performance i.e. ``defeat the adversary" \cite{li2022enhanced}. Hence we investigated the impact of resetting the critic optimizer after every training iteration i.e. after $n_d$ critic and 1 generator iterations in Algorithm~\ref{alg:WGAN_GP_training}. The results shown in Fig.~\ref{fig:nmse_iter_ro} indicate that optimizer resetting provides a $\sim 1$ dB reduction in NMSE for CDL-A for WGAN-GP ($\mathbbm{1}_{\mathrm{GP}}=1$) and a $\sim 2$ dB reduction for WGAN ($\mathbbm{1}_{\mathrm{GP}}=0$). 

\end{document}